\documentclass[reqno]{amsart}

\usepackage{amssymb,amsfonts,latexsym,amscd,epsfig }

\usepackage{graphicx}

\begin{document}

\def\A{{\mathcal A}}
\def\alg{{\mathcal A}}
\def\amp{{\rm Amp}}

\def\Bound{{\mathcal B}}

\def\bra{\big\langle}

\def\C{{\Bbb C}}
\def\cdim{{\rm codim}}
\def\crit{{\mathcal C}}

\def\Dom{\mathfrak{Dom}}
\def\det{{\rm det}}
\def\dim{{\rm dim}}
\def\dist{{\rm dist}}

\def\Exp{{\Bbb E}}
\def\Expnd {{\Bbb E}_{n-d}}
\def\Exptc{{\Bbb E}_{2-conn}}
\def\Expd{{\Bbb E}_{disc}}
\def\e{\varepsilon}
\def\en{{e_\Delta}}

\def\H{{\mathcal H}}
\def\Hess{{\rm Hess}}
\def\heta{\kappa_\eta}
\def\Hpl{{\Bbb H}}

\def\Ie{I}
\def\Im{{ Im}}

\def\Je{J_{\lambda^2}}
\def\hJe{\widehat{J}_{\lambda^2}}
\def\Jeta{J_\eta}
\def\hJeta{\widehat{J}_\eta}

\def\ket{\big\rangle}

\def\LL{\Lambda_L}
\def\LLs{\Lambda_L^*}
\def\LLd{\Lambda_{L,\frac12}}
\def\LLsd{\Lambda_{L,\frac12}^*}
\def\lb{\left[}

\def\mat{{\mathcal M}}
\def\mes{{\rm mes}}
\def\mset{K}

\def\N{{\Bbb N}}

\def\nm{{|\!|\!|\,}}

\def\p{r}
\def\pip{\tau}

\def\qm{q^{(m+1)}}

\def\R{{\Bbb R}}
\def\rb{\right]}

\def\rc{\frac{1}{2}}
\def\Rem{ R}

\def\Sc{{\mathcal S}}
\def\supp{{\rm supp}}

\def\Tor{\Bbb T}

\def\cS{\mathcal S}

\def\ualpha{\underline{\alpha}}
\def\ubeta{\underline{\beta}}
\def\up{\underline{\vp}}
\def\uq{\underline{q}}
\def\uk{\underline{\vk}}
\def\utk{\underline{\tilde\vk}}
\def\uu{\underline{u}}
\def\uvw{\underline{\vw}}
\def\uxi{\underline{\xi}}

\def\wup{{\widetilde{\underline p}}}
\def\wuq{{\widetilde{\underline q}}}
\def\wuu{{\widetilde{\underline u}}}

\def\Z{{\Bbb Z}}

\def\vx{{ x}}
\def\vy{{ y}}
\def\ve{{ e}}
\def\vk{{ k}}
\def\vl{{ l}}
\def\vm{{ m}}
\def\vn{{ n}}
\def\vp{{ p}}
\def\vQ{{ Q}}
\def\vq{{ q}}
\def\vr{{ r}}
\def\vv{{ v}}
\def\vw{{ w}}

\def\1{{\bf 1}}

\def\eqnn{\begin{eqnarray*}}
\def\eeqnn{\end{eqnarray*}}
\def\eqn{\begin{eqnarray}}
\def\eeqn{\end{eqnarray}}
\def\bal{\begin{align}}
\def\eal{\end{align}}

\def\prf{\begin{proof}}
\def\endprf{\end{proof}}



\newtheorem{theorem}{Theorem}[section]
\newtheorem{corollary}{Corollary}[section]
\newtheorem{definition}{Definition}[section]
\newtheorem{proposition}{Proposition}[section]
\newtheorem{lemma}{Lemma}[section]

\title{Convergence in higher mean of
a random Schr\"odinger to a linear
Boltzmann evolution }

\author{Thomas Chen}

\address{Department of Mathematics,
Princeton University,
Fine Hall, Washington Road,
Princeton, NJ 08544, U.S.A.}
\email{tc@math.princeton.edu}

\date{}
\maketitle

\begin{abstract}
We study the macroscopic scaling and weak coupling
limit for a random Schr\"odinger
equation on $\Z^3$.
We prove that the Wigner transforms of a large class
of "macroscopic" solutions converge in
$\p$-th mean to solutions of a linear Boltzmann equation,
for any finite value of $\p\in\R_+$.
This extends previous
results where convergence in expectation was established.
\end{abstract}

\parskip = 8 pt

\section{Introduction}

We study the macroscopic scaling and weak coupling
limit of the quantum dynamics in the three dimensional Anderson model,
generated by the Hamiltonian
\eqn
        H_\omega \; = \; -  \, \frac12 \, \Delta \, + \, \lambda  \, V_\omega(x)
        \label{eq:RSE-1}
\eeqn
on $\ell^2(\Z^3)$. Here, $\Delta$ is the nearest neighbor discrete
Laplacian, $0<\lambda\ll1$ is a small coupling constant that defines the
disorder strength, and the random potential is given by $V_\omega(x)=\omega_x$,
where $\{ \omega_x \}_{x\in\Z^3}$, are independent, identically distributed
Gaussian random variables.

While the phenomenon of impurity-induced insulation
is, for strong disorders $\lambda\gg1$
or extreme energies, mathematically well understood (Anderson localization, \cite{aimo,frsp}),
establishing the existence of electric conduction in
the weak coupling regime $\lambda\ll1$ is a key open problem of outstanding difficulty.
A particular strategy to elucidate aspects of the latter, which has led to important
recent successes (especially \cite{erdsalmyau}), is to
analyze the macroscopic transport properties derived from the microscopic
quantum dynamics generated by (\ref{eq:RSE-1}),
\cite{erd,erdyau,erdsalmyau,sp,ch}.

Let $\phi_t\in\ell^2(\Z^3)$ be the solution of the random Schr\"odinger equation
\eqn
        \left\{
        \begin{array}{rcl}
        i\partial_t\phi_t&=&H_\omega \, \phi_t \\
        \phi_0&\in&\ell^2(\Z^3) \;,
        \end{array}
        \right.
\eeqn
with a deterministic initial condition $\phi_0$
which is supported on a region of diameter $O(\lambda^{-2})$.
Let $W_{\phi_t}(x,v)$ denote its {\em Wigner transform}, where
$x\in\frac12\Z^3\equiv(\Z/2)^3$, and $v\in\Tor^3=[0,1]^3$.
We consider a scaling for small $\lambda$ defined by the
macroscopic time, position, and velocity variables
$(T,X):=\lambda^{2}(t,x)$, $V:=v$, while
$(t,x,v)$ are the microscopic variables.
Likewise, we introduce an appropriately rescaled, macroscopic counterpart
$W^{resc}_\lambda(T,X,V)$ of $W_{\phi_t}(x,v)$.

It was proved by Erd\"os and Yau for the continuum, \cite{erdyau,erd},
and by the author for the lattice model, \cite{ch}, that for any test
function $J(X,V)$, and globally in macroscopic time $T$,
\eqnn
        &&\lim_{\lambda\rightarrow0}\Exp\Big[
        \int dX \, dV \, J(X,V) \, W_\lambda^{resc}(T,X,V)\Big]
        \nonumber\\
        &&\hspace{3cm}= \; \int dX \, dV \, J(X,V) \, F(T,X,V)\;,
\eeqnn
where $F(T,X,V)$ is the solution of a linear Boltzmann equation.
For the random wave equation, a similar
result is proved by Lukkarinen and Spohn, \cite{lusp}.
The corresponding local in $T$ result was established much earlier
by Spohn, \cite{sp}.

The main goal of this paper is to improve the mode of convergence.
We establish convergence in $r$-th mean,
\eqn
        &&\lim_{\lambda\rightarrow0}\Exp\Big[\,
        \Big|\int dX \, dV \, J(X,V) \, W_\lambda^{resc}(T,X,V)
        \nonumber\\
        &&\hspace{3cm}- \, \int dX \, dV \, J(X,V) \, F(T,X,V)\Big|^\p\,\Big] \; = \; 0\;,
    \label{eq:Lr-conv-main-1}
\eeqn
for any $\p\in2\N$, and thus for any finite $\p\in\R_+$.
Thus, in particular, we observe that the variance of
$\int dX dV  J(X,V) W_\lambda^{resc}(T,X,V)$ vanishes in this macroscopic,
hydrodynamic limit.

Our proof comprises generalizations and extensions of the graph expansion methods
introduced by Erd\"os and Yau in \cite{erdyau,erd},
and further elaborated on in \cite{ch}. The structure of the graphs entering
the problem is significantly more complicated than in \cite{erdyau,erd,ch},
and the number of graphs in the expansion grows much faster than in \cite{erdyau,erd,ch}
(superfactorial versus factorial).
A main technical result in this paper
establishes that
the associated Feynman amplitudes are sufficiently small to compensate for
the large number of graphs, which is shown to imply (\ref{eq:Lr-conv-main-1}).
This is similar to the approach in \cite{erdyau,erd,ch}.

The present work addresses a time scale of order $O(\lambda^{-2})$
(as in \cite{erdyau,erd,ch}), in which the
average number of collisions experienced by the electron is finite,
so that ballistic behavior is observed. Accordingly, the
macroscopic dynamics is governed by a linear Boltzmann equation. Beyond
this time scale, the average number of collisions is {\em infinite},
and the level of difficulty of the problem increases drastically.
In their recent breakthrough result, Erd\"os,  Salmhofer and  Yau
have established that over a time scale of order
$O(\lambda^{-2-\kappa})$ for an explicit numerical value of
$\kappa>0$, the macroscopic dynamics in $d=3$ derived from the
quantum dynamics is determined by a {\em diffusion equation},
\cite{erdsalmyau}.

We note that control of the macroscopic dynamics up to a time scale
$O(\lambda^{-2})$ produces lower bounds of the same order
(up to logarithmic corrections) on the localization lengths of eigenvectors
of $H_\omega$, see \cite{ch} for $d=3$
(the same arguments are valid for $d\geq3$).
This extends recent results of Schlag, Shubin and
Wolff, \cite{shscwo}, who derived similar lower bounds for the
weakly disordered Anderson model in dimensions $d=1,2$ using
harmonic analysis techniques.

This work comprises a partial joint result with Laszlo Erd\"os
(Lemma {\ref{conngrbd-1}}), to whom the author is deeply grateful for his
support and generosity.

\section{Definition of the model and statement of main results}

To give a mathematically well-defined meaning to all quantities occurring in our analysis,
we first introduce our model on a finite box
\eqn
        \LL=\{-L,-L+1,\dots,-1,0,1,\dots,L-1,L\}^3 \; \subset \; \Z^3 \; ,
\eeqn
for $L\in\N$ much larger
than any relevant scale of the problem,
and take the limit $L\rightarrow\infty$ later.
All estimates derived in the sequel will be uniform in $L$.
We consider the discrete Schr\"odinger operator
\eqn
        H_\omega \; = \; - \, \frac12\,\Delta \, + \, \lambda V_\omega \;
        \label{Homega-def}
\eeqn
on $\ell^2(\LL)$ with periodic boundary conditions.
Here, $\Delta$ is the nearest neighbor Laplacian,
\eqn
        (\Delta f)(x) \; = \; 6f(x)-\sum_{|y-x|=1}f(y) \;,
\eeqn
and
\eqn
            V_{\omega}(\vx) \; = \; \omega_\vx
\eeqn
is a random potential with
$\{\omega_y\}_{y\in\LL}$   i.i.d. Gaussian random variables
satisfying $\Exp[\omega_x]=0$, $\Exp[\omega_x^2]=1$, for all $x\in\LL$.
Expectations of higher powers of $\omega_x$ satisfy Wick's theorem,
cf. \cite{erdyau}, and our discussion below. Clearly, $\|V_\omega\|_{\ell^\infty(\LL)}<\infty$
almost surely (a.s.), and $H_\omega$ is a.s. self-adjoint on $\ell^2(\LL)$, for every $L<\infty$.

Let $\LLs=\frac1L\Lambda_{L}=\{-1,-\frac{L-1}{L},\dots,-\frac{1}{L},0,\frac{1}{L},\dots, \frac{L-1}{L},1\}^3\subset\Tor^3$
denote the lattice dual to $\LL$, where $\Tor^3=[-1,1]^3$ the 3-dimensional unit torus.
For $0<\rho\leq1$ with $\frac1\rho\in\N$,
we define $\Lambda_{L,\rho}:= \rho\Lambda_{\rho^{-1} L}$, and note that its dual lattice
is given by $\Lambda_{L,\rho}^*=\frac{1}{ L} \Lambda_{ \rho^{-1} L}\subset \rho^{-1}\Tor^3$.
For notational convenience, we shall write $\int_{\Lambda_{L,\rho}}dk \equiv \sum_{k\in\Lambda_{L,\rho}}$,  and
$\int_{\rho^{-1}\Tor^3}dk$ for the Lebesgue integral.
For the Fourier transform and its inverse, we use the convention
\eqn
            \widehat f(\vk) \; = \; \rho^3\sum_{\vx\in \Lambda_{L,\rho}}
            e^{-2\pi i\vk\cdot\vx} f(\vx) \; \; , \; \;
            g^{\vee}(\vx)
            \; = \; \int_{\Lambda^*_{L,\rho}} dk\; g(\vk) e^{2 \pi i \vk\cdot\vx}  \;,
        \label{eq:Fourier-def-1}
\eeqn
for $L\leq\infty$ (where $\Lambda_{\infty,\rho}=\rho\Z^3$ and $\Lambda_{\infty,\rho}^*=\Tor^3/\rho$).
We will mostly use $\rho=1$, and sometimes $\rho=\frac12$.
On $\Lambda_{L,\rho}^*$, we define $\delta(k)=\widehat1(k)$ with $\delta(0)=|\Lambda_{L,\rho}|$
if $k=0$ and $\delta(k)=0$ if $k\neq0$.
On $\Tor^d$ or $\R^d$, $\delta$ will denote the usual $d$-dimensional delta distribution.
The nearest neighbor lattice Laplacian defines the Fourier multiplier
\eqn
        (-\Delta  f)\,\widehat{\;}\,(\vk) \; = \; 2 \, \en(\vk) \, \widehat f(\vk) \;,
\eeqn
where
\eqn
        \en(\vk) \; = \;  \sum_{i=1}^3 \big( 1 \, - \, \cos(2\pi k_i) \big)
        \; = \; 2 \, \sum_{i=1}^3 \sin^2(\pi k_i)
        \label{kinendef}
\eeqn
determines the kinetic energy of the electron.

Let $\phi_t\in \ell^2(\LL)$ denote the solution of the random Schr\"odinger equation
\eqn
        \left\{
        \begin{array}{rcl}
        i\partial_t\phi_t&=&H_\omega \phi_t \\
        \phi_0&\in& \ell^2(\LL)\;,
        \end{array}
        \right.
\eeqn
for a fixed realization of the random potential.
We define its (real, but not necessarily positive) Wigner transform
$W_{\phi_t}: \LLd\times\LLs\rightarrow\R$ by
\eqn
        W_{\phi_t}(x,v) \; := \; 8\sum_{y,z\in\LL \atop y+z=2x}
        \overline{\phi_t(y)} \, \phi_t(z)
        \, e^{2\pi i(y-z)v} \;.
\eeqn
Fourier transformation with respect to the variable $x\in\LLd$
(i.e. (\ref{eq:Fourier-def-1}) with $\rho=\frac12$, see  \cite{erdsalmyau} for more details) yields
\eqn
        \widehat W_{\phi_t}(\xi,v) \; = \; \overline{\widehat\phi_t(v-\frac\xi2)}
        \, \widehat\phi_t(v+\frac\xi2)\;,
        \label{FTWx}
\eeqn
for $v\in\LLs$ and $\xi\in \LLsd\subset2\Tor^3$.

The Wigner transform is the key tool in our derivation of the
macroscopic limit for the quantum dynamics described by (\ref{RSE}).
For $\eta>0$ small, we introduce macroscopic variables
$T:=\eta t$, $X:= \eta x$, $V:=v$, and consider the rescaled Wigner transform
\eqn
        W^\eta_{\phi_t}(X,V) \; := \; \eta^{-3}W_{\phi_t}(\eta^{-1}X,V)
\eeqn
for $T\geq0$, $X\in \eta \LLd$, and $V\in\LLs$.

For a Schwartz class function $J\in \Sc(\R^3\times \Tor^3)$, we write
\eqn
        \langle \, J \, , \, W^\eta_{\phi_t} \, \rangle \; := \;
        \sum_{X\in\eta\LLd}\int_{\LLs}dV \, \overline{J(X,V)}W^\eta_{\phi_t}(X,V)
        \; .
        \label{eqn3-1-0}
\eeqn
With $\widehat W_{\phi_t}$ as in (\ref{FTWx}), we have
\eqn
        \langle \, J \, , \, W^\eta_{\phi_t} \, \rangle
        \; = \;
        \langle \, \widehat J_\eta \, , \, \widehat W_{\phi_t} \, \rangle
        \; = \;
        \int_{\LLsd\times\LLs }d\xi \, dv \,
        \overline{\hJeta(\xi,v)} \,
        \widehat W_{\phi_t}(\xi,v)  \; ,
        \label{eqn3-1-1}
\eeqn
where $J_\eta(x,v):=\eta^{-3}J(\eta x,v)$, and
\eqn
        \hJeta(\xi,v) \; = \;  \eta^{-3}\sum_{x\in\LLd}J(\eta x,v)e^{-2\pi i x\xi}
    \; = \; \eta^{-3} \sum_{X\in\eta\LLd} J(X,v)e^{-\frac{2\pi i X\xi}{\eta}}
        \;.
\eeqn
We note that in the limit $L\rightarrow\infty$,
$\hJeta(\xi,v)$ tends to a smooth delta function with respect to the $\xi$-variable, of width $O(\eta)$
and amplitude $O(\eta^{-1})$,
but remains uniformly bounded with respect to $\eta$ in the $v$-variable.

The macroscopic scaling limit obtained from letting $\eta\rightarrow0$, with $\eta=\lambda^2$,
is determined by a linear Boltzmann equation.
This was proven in \cite{ch} for $\Z^3$, and non-Gaussian distributed random potentials
(the Gaussian case follows also from \cite{ch}).
The corresponding result for the
continuum model in dimensions 2, 3 was proven in \cite{erdyau}.

\begin{theorem}\label{Boltzlimthm}
For $\eta>0$, let
\eqn
        \phi_0^\eta(x) \; := \; \eta^{\frac32} \, \frac{h(\eta x)
    \, e^{2 \pi i \frac{ S(\eta x)}{\eta}}}{\|h\|_{\ell^2(\eta\Z^3)}} \;,
        \label{phi0-def}
\eeqn
with $h,S\in\cS(\R^3,\R)$ of Schwartz class, and $\|h\|_{L^2(\R^3)}=1$.
Assume
$L$ sufficiently large (see (\ref{eq:L-bd-assump-1})) that  $\phi_0^\eta\Big|_{\LL}=\phi_0^\eta$.
Let $\phi_t^\eta$ be the solution of the random
Schr\"odinger equation
\eqn
        \label{RSE}
        i \, \partial_t  \phi_t^\eta \; = \; H_\omega \,  \phi_t^\eta
\eeqn
on $\ell^2(\LL)$ with initial condition $\phi_0^\eta$, and let
\eqn
        W_T^{(\eta)}(X,V) \; := \;
        W^{\eta}_{\phi_{\eta^{-1}T}^{\eta} } (X,V)
\eeqn
denote the corresponding rescaled Wigner transform.

Choosing
\eqn
        \eta \; = \; \lambda^2 \;,
\eeqn
where $\lambda$ is the coupling constant in  (\ref{Homega-def}),
it follows that
\eqn
        \lim_{\lambda\rightarrow0}\lim_{L\rightarrow\infty}\Exp\big[ \,
        \langle \, J \, , \,  W_T^{(\lambda^2)} \, \rangle \,\big]
        \; = \; \langle \, J \, , \, F_T \, \rangle\;,
        \label{eq:W-exp-linB-1}
\eeqn
where $F_T(X,V)$ solves the linear Boltzmann equation
\eqn
    &&\partial_T F_T(X,V) \, + \, \sum_{j=1}^3 (\sin2\pi V_j)  \partial_{X_j} F_T(X,V)
    \nonumber\\
    &&\hspace{2cm}= \;  \int_{\Tor^3} dU  \, \sigma(U,V) \,
    \lb F_T(X,U) \, - \, F_T(X,V)\rb
    \label{linB}
\eeqn
with initial condition
\eqn
        F_0(X,V) & = & w-\lim_{\eta\rightarrow0}W_{\phi^\eta_0}^\eta
        \nonumber\\
        &=& |h(X)|^2 \, \delta(V \, - \, \nabla S(X))\;,
        \label{initcondweaklim}
\eeqn
and
$$
        \sigma(U,V) \; := \; 2  \pi \, \delta(\en(U) \, - \, \en(V))
$$
denotes the collision kernel.
\end{theorem}

The purpose of the present work is to obtain a significant improvement of the mode of convergence.

Our main result is the following theorem.

\begin{theorem}\label{mainthm}
Assume that the Fourier transform of (\ref{phi0-def}), $\widehat\phi_0^\eta$,
satisfies the concentration of singularity property
(\ref{eq:phi-0-dec-1}) - (\ref{eq:phi0-sing-conc-1}).
Then, for any fixed, finite
$\p\in2\N$, any $T>0$, and for any Schwartz class function $J$,
the estimate
\eqn
        \lim_{L\rightarrow\infty}\Big(\Exp\Big[ \, \Big| \, \bra \, J \, , \, W_T^{(\lambda^2)} \, \ket \, - \,
        \Exp\big[ \, \bra \, J \, , \, W_T^{(\lambda^2)} \, \ket \, \big] \, \Big|^\p\Big]
        \Big)^{\frac{1}{\p}} \; \leq \; c(r,T)
        \lambda^{\frac{1}{300\p}} \;
\eeqn
holds
for $\lambda$ sufficiently small, and a finite constant $c(r,T)$
that does not depend on $\lambda$.
Consequently,
\eqn
        \lim_{\lambda\rightarrow0}\lim_{L\rightarrow\infty}
        \Exp\Big[ \, \Big| \, \bra \, J \, , \, W_T^{(\lambda^2)} \, \ket \, - \,
        \bra \, J \, , \, F_T \, \ket \, \Big|^\p \, \Big] \; = \; 0
        \label{lpconvmainthm}
\eeqn
(i.e. convergence in $\p$-th mean),   for any finite $\p,T\in\R_+$.
\end{theorem}

We observe that, in particular, the variance
of $\bra J,W_T^{(\lambda^2)}\ket$ vanishes in the
limit $\lambda\rightarrow0$. Moreover, the following result is an immediate consequence.

\begin{corollary}
Under the assumptions of Theorem {\ref{mainthm}},
the rescaled Wigner transform $W_T^{(\lambda^2) }$ convergences weakly, and
in probability, to a solution of the linear Boltzmann equations,
globally in $T>0$, as $\lambda\rightarrow0$.
That is, for  any finite $T>0$, any $\nu>0$, and any $J$ of Schwartz class,
\eqn
        {\Bbb P}\Big[ \, \lim_{\lambda\rightarrow0}
        \Big| \, \bra J,W_T^{(\lambda^2) }\ket-\bra J,F_T\ket \, \Big| \, > \, \nu \, \Big]
        \; = \; 0 \;,
        \label{convprobmainthm}
\eeqn
where $F_T$ solves (\ref{linB}) with initial condition
(\ref{initcondweaklim}).
\end{corollary}

\subsection{Singularities of $\widehat\phi_0^\eta$}
\label{subsec:phi-0-sing-1}

One obtains a well-defined semiclassical initial condition
(\ref{initcondweaklim}) for the linear Boltzmann evolution (\ref{linB})
if the initial condition is of WKB type (\ref{phi0-def}),
but in general not if the initial condition is
only required to be in $\ell^2(\Z^3)$.
However, for
the expected value of the quantum fluctuations
in (\ref{eq:W-exp-linB-1}) to converge to zero as
$\lambda\rightarrow0$, it is sufficient to have initial
data in $\ell^2(\Z^3)$, see \cite{erdyau,ch}.

As we will see, a key point in proving that as $\lambda\rightarrow0$,
the quantum fluctuations vanish
in higher mean, i.e. (\ref{lpconvmainthm}),
it is necessary to control the overlap of the singularities of $\widehat\phi_0^\eta$
with those of the resolvent multipliers  $(\en(k)-\alpha \pm i\e)^{-1}$,
where $\alpha\in\R$ and $\e=O(\eta)\ll1$.
As opposed to the case in (\ref{eq:W-exp-linB-1}),
it cannot be expected that the quantum fluctuations vanish in higher mean for general $L^2$
inital data (for (\ref{eq:W-exp-linB-1}), the overlap of the singularities of
$\widehat\phi_0^\eta$ and of those of the resolvent multipliers plays no r\^ole).
Moreover, we note that the singularities of the WKB initial condition
\eqn
        \widehat \phi_0^\eta(k) & = &
        \eta^{\frac32} \sum_{x\in \Z^3} h(\eta x) \, e^{2 \pi i (\frac{S(\eta x)}{\eta}-kx)}
        \nonumber\\
        & = & \eta^{\frac32}
    \sum_{X\in \eta\Z^3} h(X) \, e^{2 \pi i (\frac{S(X)-kX}{\eta})} \;
\eeqn
(which are determined by the zeros of $\det \, \Hess \, S(X)$,
the determinant of the Hessian of $S$)
will possess a rather arbitrary structure for generic choices of $S\in\cS(\R^3,\R)$.
At present, we do not know if
for WKB initial data of the form (\ref{phi0-def}),
the quantum fluctuations would converge
to zero in higher mean without any further restrictions on
the phase function $S\in\cS(\R^3,\R)$.
A more detailed analysis of these questions is left for future work.
In this paper, we shall assume
that the Fourier transform of the WKB initial condition (\ref{phi0-def})
satisfies a {\em concentration of singularity condition}:
\eqn
        \widehat\phi_0^\eta(k) \; = \; f_{\infty}^\eta(k) \, + \, f_{sing}^\eta(k)  \;,
        \label{eq:phi-0-dec-1}
\eeqn
where
\eqn
            \| \, f_{\infty}^\eta  \, \|_{L^\infty(\Tor^3)} \; < \; c   \;,
        \label{eq:phi0-sing-conc-0}
\eeqn
and
\eqn
        \| \, |f_{sing}^\eta|*|f_{sing}^\eta|\, \|_{L^2(\Tor^3)}
        \; = \;
        \|\,|f_{sing}^\eta|^\vee\,\|_{\ell^4(\Z^3)}^2
        \; \leq \; c'\,\eta^{\frac45}
        \label{eq:phi0-sing-conc-1}
\eeqn
for finite, positive constants $c$, $c'$ independent of $\eta$.
This condition imposes a restriction on the possible choices of the phase function $S$.

The following simple, but physically important examples of
$\widehat \phi_0^\eta$ satisfy (\ref{eq:phi-0-dec-1}) - (\ref{eq:phi0-sing-conc-1}).

\subsubsection{Example}

Let $S(X)=p X$ for $X\in\supp\{\,h\,\}$, and $p\in\Tor^3$. Then,
\eqn
        \widehat\phi_0^\eta(k) \; = \; \frac{\eta^{-\frac32}
        \widehat h ( \eta^{-1}(k-p))}{\|h\|_{\ell^2(\eta\Z^3)}}
         \; =: \;   \delta_\eta(k-p) \;.
         \label{eq:osc-sum-bump-1}
\eeqn
Since $h$ is of Schwartz class, $\delta_\eta$ is a smooth bump function
concentrated on a ball of radius $O(\eta)$, with $\|\delta_\eta\|_{L^2(\Tor^3)}=1$.
Accordingly, we find
\eqn
        (|\delta_\eta|*|\delta_\eta|)(k) \; \approx \; \chi(|k|<c\eta) \;,
\eeqn
and
\eqn
        \| \, |\delta_\eta|*|\delta_\eta| \, \|_{L^2(\Tor^3)}
        \; = \;
        \| \, |\delta_\eta|^\vee \, \|_{\ell^4(\Z^3)}^2
        \; \leq \; c \, \eta^{\frac32} \;.
\eeqn
Hence,
(\ref{eq:phi-0-dec-1}) - (\ref{eq:phi0-sing-conc-1}) is satisfied,  with $f_\infty^\eta=0$.
We remark that in this example, $p\in\Tor^3$ corresponds to the velocity of the macroscopic
initial condition $F_0(X,V)$ in (\ref{initcondweaklim}) for the linear Boltzmann evolution.

\subsubsection{Example}

As a small generalization of the previous case, we may likewise assume for
$S$ that for every $k\in\Tor^3$, there are finitely many solutions $X_j(k)$
of $\nabla_X S(X_j(k))=k$, and that $X_j(\,\cdot\,)\in C^1(\supp\{\,h\,\})$
for each $j$. Moreover, we assume that
$|\det \, \Hess \, S(X)| \, > \, c$ uniformly on $\supp\{ \, h \, \}$.
Then, by stationary phase arguments, \cite{st}, one finds that
\eqn
        \widehat \phi_0^\eta(k) \; = \; f_\infty^\eta(k) \, +  \, f_{sing}^\eta(k)
        \; \; \; \; , \; \; \; \;
        \|f_{\infty}^\eta\|_{L^\infty(\Tor^3)} \; < \; c
\eeqn
with
\eqn
         f_{sing}^\eta(k) \; = \; \sum_{j }  c_j \, \delta_\eta^{(j)}(k-\nabla_X S(X_j(k))) \;,
\eeqn
for constants $c_j$ independent of $\eta$, and smooth bump functions $\delta_\eta^{(j)}$
similar to (\ref{eq:osc-sum-bump-1}).  One again obtains
$\|\,|f_{sing}^{\eta}|^{\vee}\,\|_{\ell^4(\Z^3)}^2\leq c\eta^{\frac32}$, which verifies that
(\ref{eq:phi-0-dec-1}) - (\ref{eq:phi0-sing-conc-1}) holds.
$\nabla S$ determines the velocity distribution of the macroscopic
initial condition $F_0(X,V)$ in (\ref{initcondweaklim}).

\section{Proof of Theorem {\ref{mainthm}}}

We expand $\phi_t$ into a truncated Duhamel series
\eqn
        \phi_t \; = \; \sum_{n=0}^{N-1} \phi_{n,t} \, + \, R_{N,t} \;,
        \label{eq:Duham-exp-def-1}
\eeqn
where
\eqn
        \phi_{n,t} \; := \; (-i\lambda)^n \int_{\R_+^{n+1}} ds_0\cdots ds_n \,
        \delta(\sum_{j=0}^n s_j-t)
        \,e^{i s_0 \frac\Delta2} \, V_\omega \, e^{is_1\frac\Delta2}
        \cdots V_\omega \, e^{is_n\frac\Delta2} \, \phi_0
        \;\;\;\;
\eeqn
denotes the $n$-th Duhamel term, and where
\eqn
        R_{N,t} \; = \; - \, i\lambda\int_0^t ds \, e^{-i(t-s)H_\omega} \, V_\omega \, \phi_{N-1,s} \;
\eeqn
is the remainder term. Here and in the sequel, we write $\phi_0\equiv\phi_0^\eta$ for
brevity. The number $N$ remains to be optimized.
Since $\|V_\omega\|_{\ell^1(\LL)}<\infty$ a.s., $\widehat V_\omega$  is well-defined
and bounded on $\LLs$, with probability one, for every $L<\infty$. Then,
\eqn
        \widehat \phi_{n,t}(k_0)&=&(-i\lambda)^n\int ds_0\cdots ds_n
        \, \delta(\sum_{j=0}^n s_j-t)
        \nonumber\\
        && \,\int_{(\LLs)^n}dk_1\cdots
        dk_n \, e^{-is_0 \en(k_0)} \, \widehat V_\omega(k_1-k_0) \, e^{-is_1 \en(k_1)} \,
        \cdots
        \nonumber\\
        &&\hspace{2.5cm}
        \cdots \, \widehat V_\omega(k_{n}-k_{n-1}) \,
        e^{-is_n \en(k_n)} \, \widehat \phi_0(k_n) \;.
\eeqn
Expressed as a resolvent expansion in momentum space, we find
\eqn
        \widehat \phi_{n,t}(k_0)&=&\frac{(-\lambda)^n}{2\pi i} \, e^{\e t} \, \int_{\R}  d\alpha \,
        e^{-it\alpha}
        \nonumber\\
        && \,\int_{(\LLs)^n}dk_1\cdots
        dk_n \, \frac{1}{\en(k_0)-\alpha-i\e} \, \widehat V_\omega(k_1-k_0)
        \nonumber\\
        &&\hspace{2 cm}
        \cdots \, \widehat V_\omega(k_{n}-k_{n-1}) \, \frac{1}{\en(k_n)-\alpha-i\e} \,
        \widehat \phi_0(k_n) \;.
        \label{hatphint-expans}
\eeqn
We refer to the Fourier multiplier $\frac{1}{\en(k)-\alpha-i\e}$ as a
{\em particle propagator}.
Likewise, we note that (\ref{hatphint-expans}) is
equivalent to the $n$-th term in the resolvent expansion of
\eqn
        \phi_t \; = \; \frac{1}{2\pi i}\int_{-i\e+\R} \, dz \, e^{-itz} \, \frac{1}{H_\omega-z} \, \phi_0
        \;.
\eeqn
By the analyticity of the integrand in (\ref{hatphint-expans})
with respect to the variable $\alpha$,
the path of the $\alpha$-integration
can, for any fixed $n\in\N$, be deformed
into the closed contour
\eqn
        \Ie \; = \; I_0 \, \cup \, I_1 \;,
        \label{defIloop}
\eeqn
away from $\R$, with
\eqnn
        I_0 &:=& [-1, 13]\\
        I_1&:=& ([-1, 13]-i)\cup (-1-i(0,1]) \cup (13-i(0,1])   \;,
\eeqnn
which encloses ${\rm spec}\big(- \Delta -i\e\big) = [0,12]-i\e $.

Next, we apply the time partitioning method introduced in
\cite{erdyau}. To this end, we choose $\kappa\in\N$ with
$1\ll\kappa\ll\e^{-1}$, and subdivide $[0,t]$ into $\kappa$ subintervals
bounded by $\theta_j=\frac{jt}{\kappa}$, $j=1,\dots,\kappa$.
Then,
\eqn
        R_{N,t} \; = \; - \, i\lambda \, \sum_{j=0}^{\kappa-1}e^{-i(t-\theta_{j+1})H_\omega}
        \int_{\theta_j}^{\theta_{j+1}} ds \, e^{-i(\theta_{j+1}-s)H_\omega}
        \, V_\omega \, \phi_{N-1,s} \;.
            \label{RemNt-def-1}
\eeqn
Let $\phi_{n,N,\theta}(s)$ denote the $n$-th Duhamel term, conditioned on
the requirement that the first $N$ collisions occur
in the time interval $[0,\theta]$, and all remaining $n-N$ collisions
take place in the time interval $(\theta,s]$.
That is,
\eqn
        \phi_{n,N,\theta}(s)&:=&(-i\lambda)^{n-N}
        \int_{\R_+^{n-N+1}} ds_{0}\cdots ds_{n-N} \,
        \delta(\sum_{j=0}^{n-N}s_j-(s-\theta))
        \nonumber\\
        &&\hspace{2cm}\,
        e^{is_0 \frac\Delta2} \, V_\omega \, \cdots \, V_\omega \, e^{is_{n-N}\frac\Delta2} \,
        V_\omega \, \phi_{N-1,\theta}\;.
\eeqn
Moreover, let
\eqn
        \widetilde\phi_{n,N,\theta}(s) \; := \;  - \, i\lambda   \,
        V_\omega \, \phi_{n-1,N,\theta}(s)
        \label{eq:tilde-phi-def-1}
\eeqn
denote its "truncated" counterpart.
Further expanding $e^{-isH_\omega}$ in (\ref{RemNt-def-1})
into a truncated Duhamel series with $3N$ terms, we find
\eqn
        R_{N,t} \; = \; R_{N,t}^{(<4N)} \, + \, R_{N,t}^{(4N)} \; ,
\eeqn
where
\eqn
        R_{N,t}^{(<4N)} \; = \;
        \sum_{j=1}^{\kappa} \, \sum_{n=N}^{4N-1} \,
        e^{-i(t-\theta_j)H_\omega} \, \phi_{n,N,\theta_{j-1}}(\theta_{j})
        \label{eq:Rn-4N-1}
\eeqn
and
\eqn
        R_{N,t}^{(4N)} \; = \; \sum_{j=1}^{\kappa}e^{-i(t-\theta_j)H_\omega}
        \int_{\theta_{j-1}}^{\theta_j}ds \;
        e^{-i(\theta_j-s)H_\omega}
        \widetilde\phi_{4N,N,\theta_{j-1}}(s) \;.
        \label{eq:Rn-4N-2}
\eeqn
By the Schwarz inequality,
\eqn
        \|R_{N,t}^{(<4N)}\|_2 \; \leq \; 3 \, N \, \kappa \, \sup_{N\leq n<4N,1\leq j\leq\kappa}
        \|\phi_{n,N,\theta_{j-1}}(\theta_{j})\|_2
\eeqn
and
\eqn
        \|R_{N,t}^{(4N)}\|_2 \; \leq \; t \sup_{1\leq j\leq\kappa}
        \sup_{s\in[\theta_{j-1},\theta_j]}
        \|\widetilde\phi_{4N,N,\theta_{j-1}}(s)\|_2 \;,
\eeqn
for every fixed realization of $V_\omega$.

Let $\p\in 2\N$, and let
\eqn
        W_{t;n_1,n_2}(x,v) \; := \; 8\sum_{y,z \in\LL\atop y+z=2x} \, \overline{\psi_{n_2,t}(y)} \,
        \psi_{n_1,t}(z) \,
        e^{2\pi i (y-z)\cdot v} \;,
\eeqn
for $x\in \LLd$, denote the $(n_1,n_2)$-th term in the
Wigner distribution, with
\eqn
        \psi_{n,t} \; := \; \left\{\begin{array}{ll}
        \phi_{n,t}&{\rm if}\;n < N\\
        \sum_{j=1}^{\kappa} \,
        e^{-i(t-\theta_j)H_\omega} \, \phi_{n,N,\theta_{j-1}}(\theta_{j})&{\rm if} \; N\leq n<4N\\
        R_{N,t}^{(4N)}&{\rm if}\;n=4N\;.
        \end{array}\right.
        \label{psidef}
\eeqn
We note that Fourier transformation with respect to $x\in\LLd$ (see
(\ref{eq:Fourier-def-1})) yields
\eqn
        \widehat W_{t;n_1,n_2}(\xi,v) \; = \; \overline{\widehat\psi_{n_2,t}(v-\frac \xi2)} \,
        \widehat\psi_{n_1,t}(v+\frac\xi2)\;,
\eeqn
see also (\ref{FTWx}).
Then, clearly,
\eqn
        &&\Big(\Exp\Big[\big(\langle \, \hJe \, , \, \widehat W_{\phi_t} \, \rangle
        -\Exp\langle \, \hJe \, , \, \widehat W_{\phi_t} \, \rangle\big)^\p\Big]\Big)^{\frac1\p}
        \nonumber\\
        &&\hspace{1cm}\leq \; C \, N \, \sum_{n_1,n_2=0}^{4N}
        \Big(\Exp\Big[\big|\langle \, \hJe \, , \, \widehat W_{t;n_1,n_2} \, \rangle
        -\Exp\langle \, \hJe \, , \, \widehat W_{t;n_1,n_2} \, \rangle\big|^\p\Big]\Big)^{\frac1\p}\;,
    \label{eq:Exp-Lr-basic-est-1}
\eeqn
and we distinguish the following cases.

If $n_1,n_2 < N$, we note that
\eqn
        &&\Big(\Exp\Big[\big|\langle \, \hJe \, , \, \widehat W_{t;n_1,n_2} \, \rangle
        -\Exp\langle \, \hJe \, , \, \widehat W_{t;n_1,n_2} \, \rangle\big|^\p\Big]\Big)^{\frac1\p}
        \nonumber\\
        &&\hspace{1cm}
        =\Big(\Exptc \Big[\big|\langle \hJe,\widehat W_{t;n_1,n_2}
        \rangle \big|^\p\Big]\Big)^{\frac1\p}
        \nonumber\\
        &&\hspace{1cm}=\Big(\Exptc \Big[ \Big(\Big|\int d\xi dv \hJe(\xi, v) \,
        \overline{\phi_{n_2,t}(v-\frac\xi2)} \,
        \phi_{n_1,t}(v+\frac\xi2)
        \Big|^2\Big)^{\frac\p2}  \Big]\Big)^{\frac1\p}\;,
        \;\;\;\;\;\;
\eeqn
where $\Exptc $ denotes the expectation based on {\em 2-connected graphs},
cf. Definition {\ref{Expnd-def}} below.

If $N\leq n_i\leq 4N$ for at least one value of $i$, we use
\eqn
        |\langle \, \hJe \, , \, \widehat W_{t;n_1,n_2} \, \rangle|&=& \Big|\int d\xi \, d v \,
        \hJe(\xi,v) \, \overline{\psi_{n_2,t}(v-\frac\xi2)} \,
        \psi_{n_1,t}(v+\frac\xi2)\Big|
        \nonumber\\
        &\leq&\Big( \int d\xi \, \sup_{v}|\hJe(\xi,v)|\Big)
        \|\psi_{n_1,t}\|_2\|\psi_{n_2,t}\|_2
\eeqn
and
\eqn
    \int_{2\Tor^3} d\xi \, \sup_v|\hJe(\xi,v) | \; < \; c \;.
    \label{eq:hJe-int-bd-1}
\eeqn
Then, for constants $C$ which are independent of $\e$, we obtain the following
estimates.

If $n_{1} < N$ and $N\leq n_{2}< 4N$,
the Schwarz inequality implies
\eqn
        \lefteqn{
        \Big(\Exp\Big[\big|\langle \, \hJe \, , \, \widehat W_{t;n_1,n_2} \, \rangle
        \, - \, \Exp\langle \, \hJe \, , \, \widehat W_{t;n_1,n_2} \,
        \rangle\big|^{\p}\Big]\Big)^{\frac{1}{\p}}
        }
        \nonumber\\
        &\leq& C\Big\{
        \Big(\Exp\Big[\|\psi_{n_1,t}\|_2^\p\|\psi_{n_2,t}\|_2^\p \Big]
        \Big)^{\frac{1}{\p}}
        \, + \,
        \Exp\Big[  \|\psi_{n_1,t}\|_2\|\psi_{n_2,t}\|_2\Big]\Big\}
        \\
        &\leq& C \, \Big\{
        \Big(\Exp\Big[\|\psi_{n_1,t}\|_2^{2\p}\Big]\Big)^{\frac{1}{2\p}}
        \Big(\Exp\Big[\|\psi_{n_2,t}\|_2^{2\p} \Big]
        \Big)^{\frac{1}{2\p}} \,
        + \,
        \Big(\Exp\Big[  \|\psi_{n_1,t}\|_2^2\Big]\Exp\Big[\|\psi_{n_2,t}\|_2^2\Big]
        \Big)^{\frac12}\Big\}
        \;.
        \nonumber
\eeqn
Thus, if $n_1 < N$, $N \leq n_2<4N$,
\eqn
        \lefteqn{
        \Big(\Exp\Big[\big|\langle \, \hJe \, , \, \widehat W_{t;n_1,n_2} \, \rangle
        \, - \, \Exp\langle \, \hJe \, , \, \widehat W_{t;n_1,n_2} \, \rangle\big|^{\p}\Big]\Big)^{\frac{1}{\p}}
        }
        \nonumber\\
        &\leq & C \, \kappa \, N \, \Big\{\sup_{j}
        \Big(\Exp\Big[\|\phi_{n_1,t}\|_2^{2\p}\Big]\Big)^{\frac{1}{2\p}}
        \Big(\Exp\Big[\|\phi_{n_2,N,\theta_{j-1}}(\theta_j)
        \|_2^{2\p} \Big]\Big)^{\frac{1}{2\p}}
        \nonumber\\
        &&\hspace{2cm}+ \,
        \sup_{j} \Big(\Exp\Big[  \|\phi_{n_1,t}\|_2^2\Big]
        \Exp\Big[\|\phi_{n_2,N,\theta_{j-1}}(\theta_j)\|_2^2\Big]\Big)^{\frac12}\Big\}
        \;.
\eeqn
while for $n_1 < N$, $n_2=4N$,
\eqn
        \lefteqn{
        \Big(\Exp\Big[\big|\langle \, \hJe \, , \, \widehat W_{t;n_1,4N} \, \rangle
        \, - \, \Exp\langle \, \hJe \, , \, \widehat W_{t;n_1,4N} \, \rangle\big|^{\p}\Big]\Big)^{\frac{1}{\p}}
        }
        \nonumber\\
        &\leq & C \, t \, \Big\{\sup_{j}\sup_{s\in[\theta_{j-1},\theta_j]}
        \Big(\Exp\Big[\|\phi_{n_1,t}\|_2^{2\p}\Big]\Big)^{\frac{1}{2\p}}
        \Big(\Exp\Big[\|\widetilde\phi_{4N,N,\theta_{j-1}}(s)\|_2^{2\p} \Big]\Big)^{\frac{1}{2\p}}
        \\
        &&\hspace{2cm}+ \,
        \sup_{j}\sup_{s\in[\theta_{j-1},\theta_j]} \Big(\Exp\Big[  \|\phi_{n_1,t}\|_2^2\Big]
        \Exp\Big[\|\widetilde\phi_{4N,N,\theta_{j-1}}(s)\|_2^2\Big]\Big)^{\frac12}\Big\}
        \;. \nonumber
\eeqn
If $N\leq n_1,n_2\leq 4N$, we use the Schwarz inequality in the form
\eqn
        \lefteqn{
        \Big(\Exp\Big[\big|\langle \, \hJe \, , \, \widehat W_{t;n_1,n_2} \, \rangle
        \, - \, \Exp\langle \, \hJe \, , \, \widehat W_{t;n_1,n_2} \,
        \rangle\big|^{\p}\Big]\Big)^{\frac{1}{\p}}
        }
        \nonumber\\
        &\leq& C\Big\{
        \Big(\Exp\Big[\Big(\|\psi_{n_1,t}\|_2+\|\psi_{n_2,t}\|_2\Big)^{2\p} \Big]
        \Big)^{\frac{1}{\p}}
        \, + \,
        \Exp\Big[  \|\psi_{n_1,t}\|_2\|\psi_{n_2,t}\|_2\Big]\Big\}
        \nonumber\\
        &\leq& C\sum_{j=1}^2\Big\{
        \Big(\Exp\Big[\|\psi_{n_j,t}\|_2^{2\p}\Big]\Big)^{\frac{1}{\p}}
        \, + \,
        \Exp\Big[  \|\psi_{n_j,t}\|_2^2\Big]
        \Big\}
        \;.
\eeqn
Hence, for $N\leq n_1, n_2 \leq 4N$,
\eqn
        \lefteqn{
        \sum_{N\leq n_1, n_2 \leq 4N}
        \Big(\Exp\Big[\big|\langle \, \hJe \, , \, \widehat W_{t;n_1,n_2} \, \rangle
        \, - \, \Exp\langle \, \hJe \, , \, \widehat W_{t;n_1,n_2} \, \rangle\big|^{\p}\Big]\Big)^{\frac{1}{\p}}
        }
        \nonumber\\
        &\leq& C \, (N\kappa)^{2} \,
        \Big\{\sup_j\sup_{N\leq n<4N}
        \Big(\Exp\Big[\|
        \phi_{n,N,\theta_{j-1}}(\theta_j)\|_2^{2\p}\Big)^{\frac{1}{\p}}
        \nonumber\\
        &&\hspace{4cm}
        \, + \, \sup_j \sup_{N\leq n<4N}
        \Exp\Big[  \| \phi_{n,N,\theta_{j-1}}(\theta_j)\|_2^2\Big]
        \Big\}
        \nonumber\\
        &&\, + \, C \, t^2 \,
        \Big\{\sup_{j}\sup_{s\in[\theta_{j-1},\theta_j]}
        \Big(\Exp\Big[\| \widetilde\phi_{4N,N,\theta_{j-1}}(s)\|_2^{2\p} \Big]
        \Big)^{\frac{1}{\p}}
        \\
        &&\hspace{4cm}+ \, \sup_j \sup_{s\in[\theta_{j-1},\theta_j]}
        \Exp\Big[\| \widetilde\phi_{4N,N,\theta_{j-1}}(s)\|_2^2\Big]  \Big\}
        \;. \nonumber
\eeqn
We shall next use Lemmata  {\ref{mainlm1}}, {\ref{mainlm2}},
and {\ref{mainlm3}} below to bound the above sums.

From Lemma {\ref{mainlm1}},  and $((n\p)!)^{\frac{1}{\p}}<n^n \p^n$, one
obtains
\eqn
        &&
        \sum_{n_1,n_2 < N}
        \Big(\Exptc \Big[\big|\langle \,  \hJe \, , \, \widehat W_{t;n_1,n_2} \, \rangle
        \big|^\p\Big]\Big)^{\frac1\p}
        \nonumber\\
        &&\hspace{4cm}\leq \;
        C \, \e^{\frac{1}{5\p}} \, N^{N+2} \,
        (\log\frac1\e)^{3} \,
        (c\p\lambda^2\e^{-1}\log\frac1\e)^{N}\;.
        \label{keyintest-1}
\eeqn
From (\ref{aprioribd-1}) and Lemma {\ref{mainlm2}},
\eqn
        \lefteqn{
        \Big(\sum_{n_1 < N\atop N \leq n_2<4N}+\sum_{n_2 < N\atop N\leq n_1<4N}\Big)
        \Big(\Exp\Big[\big|\langle \, \hJe \, , \, \widehat W_{t;n_1,n_2} \, \rangle
        \, - \, \Exp\langle \hJe,\widehat W_{t;n_1,n_2}\rangle\big|^{\p}\Big]\Big)^{\frac{1}{\p}}
        }
        \nonumber\\
        &\leq& \Big[  C \, \kappa \, N^{N+3} \,
        (\log\frac1\e)^{3} \,
        (c\p\lambda^2\e^{-1}\log\frac1\e)^{N} \,
        \label{keyintest-2}\\
        &&\hspace{1cm}
        \Big[\frac{(c\lambda^2\e^{-1})^{4N}}{\sqrt{N!}} \, + \, (4N)^{4N} \,
        \Big( \e^{\frac{1}{5}} \, + \,
        \e^{\frac{1}{5\p}}\Big) \,
        (\log\frac1\e)^{3} \,
        (c\p\lambda^2\e^{-1}\log\frac1\e)^{4N}\Big]\Big]^{\frac12}
        \;.
        \nonumber
\eeqn
From (\ref{aprioribd-1}) and Lemma  {\ref{mainlm3}},
\eqn
        && \Big(\sum_{n_1 < N\atop n_2=4N} \, + \, \sum_{n_2 < N\atop n_1=4N}\Big)
        \Big(\Exp\Big[\big|\langle \, \hJe \, , \, \widehat W_{t;n_1,4N} \, \rangle
        \, - \, \Exp\langle \hJe,\widehat W_{t;n_1,4N}\rangle\big|^{\p}\Big]\Big)^{\frac{1}{\p}}
        \nonumber\\
        &&\hspace{1cm}\leq \; \Big( C \, \e^{-1} \,  N^{5N+1} \, \kappa^{-2N} \,
        (\log\frac1\e)^{6} \,
        (c\p\lambda^2\e^{-1}\log\frac1\e)^{5N} \Big)^{\frac12}
        \;.
        \label{keyintest-3}
\eeqn
Finally, from Lemmata {\ref{mainlm2}} and {\ref{mainlm3}},
\eqn
        \lefteqn{
         \sum_{N\leq n_1, n_2 \leq 4N}
        \Big(\Exp\Big[\big|\langle \, \hJe \, , \, \widehat W_{t;n_1,n_2} \, \rangle
        \, - \, \Exp\langle \hJe,\widehat W_{t;n_1,n_2}\rangle\big|^{\p}
        \Big]\Big)^{\frac{1}{\p}}
        }
        \nonumber\\
        &\leq&  C \, (N\kappa)^{2} \,
        \frac{(c\lambda^2\e^{-1})^{4N}}{\sqrt{N!}}
        \nonumber\\
        &&\hspace{1.5cm}
        \, + \, C \, (N\kappa)^{2}(4N)^{4N}
        \Big( \e^{\frac{1}{5}} \, + \,
        \e^{\frac{1}{5\p}}\Big) \,
        (\log\frac1\e)^{3} \,
        (c \p\lambda^2\e^{-1}\log\frac1\e)^{4N}
        \nonumber\\
        &&\hspace{1.5cm}+ \, C \, \e^{-2} \, \kappa^{-2N} \,
        (4N)^{4N} \,
        (\log\frac1\e)^{3} \,
        (c\p\lambda^2\e^{-1}\log\frac1\e)^{4N}
        \;,
        \label{keyintest-4}
\eeqn
We emphasize that the bounds (\ref{keyintest-1}) - (\ref{keyintest-4}) are {\em uniform} in $L$.

Consequently, for a choice of parameters
\eqn
        \e&=&\frac1t=\frac{\lambda^2}{T}
        \nonumber\\
        N&=&\left\lfloor\frac{\log\frac1\e}{100 \, \p \, \log\log\frac1\e}\right\rfloor
        \nonumber\\
        \kappa&=&\left\lceil(\log\frac1\e)^{150r}\right\rceil \;,
        \label{eq:Param-choice-1}
\eeqn
we find, for sufficiently small $\e$,
\eqn
        \e^{-\frac{1}{70r}}\;\;\;<\;\;\;N^N&<&\e^{-\frac{1}{100r}}
        \nonumber\\
        (4N)^{4N}&<&\e^{-\frac{1}{20r}}
        \nonumber\\
        (c\p\lambda^2\e^{-1}\log\frac1\e)^{4N}&<& \e^{-\frac{1}{50r}}
        \nonumber\\
        \frac{(c\lambda^2\e^{-1})^{4N}}{\sqrt{N!}}&<&\e^{\frac{1}{25r}}
        \nonumber\\
        \kappa^{-2N}&\leq&\e^{3}\;,
\eeqn
whereby it is easy to verify that
\eqn
        (\ref{keyintest-1})&<&
        (\log\frac1\e)^{10}\e^{\frac{1}{5r}} \, \e^{-\frac{1}{100r}}
        \e^{\frac{1}{50r}}
        \; < \; \e^{\frac{1}{20r}}
        \nonumber\\
        (\ref{keyintest-2})&<&\Big((\log\frac1\e)^{10+150r}\e^{-\frac{1}{50r}}
        \Big(\e^{\frac{1}{25r}} \, + \, \e^{-\frac{1}{20r}}
        (\e^{\frac{1}{5}}+\e^{\frac{1}{5\p}})\e^{-\frac{1}{50r}}\Big)\Big)^{\frac12}
        \; < \; \e^{\frac{1}{120r}}
        \nonumber\\
        (\ref{keyintest-3})&<&\Big((\log\frac1\e)^{10}\e^{-1-\frac{1}{20r}}
        \e^3\e^{-\frac{1}{50r}}\Big)^{\frac12}
        \; < \; \e^{\frac12}
        \nonumber\\
        (\ref{keyintest-4})&<&
        (\log\frac1\e)^{2+300r}\e^{\frac{1}{25r}}
        \, + \, (\log\frac1\e)^{6+300r} \e^{-\frac{1}{20r}}
        (\e^{\frac{1}{5}}+\e^{\frac{1}{5\p}})\e^{-\frac{1}{50r}}
        \nonumber\\
        &&
        + \, (\log\frac1\e)^3\e^{-2}\e^3\e^{-\frac{1}{20r}}\e^{-\frac{1}{50r}}
        \; < \; \e^{\frac{1}{30r}}\;.
\eeqn
Collecting all of the above, and recalling (\ref{eq:Exp-Lr-basic-est-1}),
\eqn
        \Big(\Exp\Big[\big|\langle \hJe,\widehat W_{\phi_t}\rangle
        -\Exp\langle \hJe,\widehat W_{\phi_t}\rangle\big|^\p\Big]\Big)^{\frac1\p}
        \; < \; C \, N \,\e^{\frac{1}{120\p}} \; < \; \e^{\frac{1}{150 \p}}\;,
        \label{mainvarest}
\eeqn
uniformly in $L$.
Hence, using Theorem {\ref{Boltzlimthm}},
\eqn
        \lefteqn{
        \lim_{\lambda \rightarrow0 }\lim_{L\rightarrow\infty}
        \Big(\Exp\Big[\big(\langle  J, W_{T}^{(\lambda^2)}\rangle
        - \langle  J ,F_T\rangle\big)^\p\Big]\Big)^{\frac1\p}
        }
        \nonumber\\
        &&\leq \;
        \lim_{\lambda\rightarrow0 }\lim_{L\rightarrow\infty}
        \Big(\Exp\Big[\big(\langle \hJe,\widehat W_{\phi_{t}}\rangle
        -\Exp\langle \hJe,\widehat W_{\phi_{t}}\rangle\big)^\p\Big]\Big)^{\frac1\p}
        \nonumber\\
        &&\hspace{3cm}
        + \, \lim_{\lambda\rightarrow0}\lim_{L\rightarrow\infty}
        \Big|\Exp\langle J, W_{T}^{(\lambda^2)}\rangle
        -\langle J,F_T\rangle\Big|
        \; = \; 0 \;,
        \label{mainvarest-1}
\eeqn
for every fixed, finite value of $\p\in2\N$ and $T>0$. This in turn
implies that (\ref{mainvarest-1}) holds for any fixed,
finite $\p\in\R_+$ and globally in $T$, which establishes
Theorem {\ref{mainthm}}.

\section{Main Lemmata}

In this section, we summarize the key technical lemmata needed to establish (\ref{mainvarest}).
The proofs are based on graph expansion techniques
and estimation of high dimensional singular integrals in momentum space.
To arrive at our results, we have to significantly generalize and extend methods
developed in \cite{erdyau} and \cite{ch}. In all that follows, we suppose that $L$ is finite, but
much larger than any relevant scale of the problem; for our purposes, the assumption that
\eqn
        L \; \gg \; \e^{-\p/\e}
        \label{eq:L-bd-assump-1}
\eeqn
will suffice.

\begin{lemma}
\label{mainlm1}
Let $\bar n:= n_1+n_2$, where $n_1,n_2 < N$.
For any fixed $\p \in2\N$,
and every $T=\lambda^2\e^{-1}>0$, there exists a finite constant $c=c(T)$ independent of $L$
such that
\eqn
        \Big(\Exptc \Big[\big|\langle \, \hJe \, , \, \widehat W_{t;n_1,n_2} \, \rangle
        \big|^\p\Big]\Big)^{\frac1\p}
        \; \leq \;
        \e^{\frac{1}{5r}} \, ( (\frac{\bar n \p}{2} ) !)^{\frac{1}{\p}} \,
        (\log\frac1\e)^{3} \,
        (c\lambda^2\e^{-1}\log\frac1\e)^{\frac{\bar n}{2}} \;.
        \label{keyest-1}
\eeqn
Furthermore, for any fixed $\p\in2\N$ and $n < N$,
there is an a priori bound
\eqn
        \Big(\Exp\Big[ \|\phi_{n,t}\|_2^{2\p}
        \Big]\Big)^{\frac1\p}
        \; \leq \;
        ((n\p)!)^{\frac{1}{\p}}(\log\frac1\e)^{3}
        (c\lambda^2\e^{-1}\log\frac1\e)^{n} \;,
        \label{aprioribd-1}
\eeqn
where $c$ is independent of $T$ and $L$.
\end{lemma}

The gain of a factor $\e^{\frac{1}{5\p}}$ in (\ref{keyest-1}) over
the a priori bound (\ref{aprioribd-1}) is the key ingredient in our proof
of (\ref{mainvarest}).

\begin{lemma}
\label{mainlm2}
For any fixed $\p\in2\N$,  $N\leq n<4N$, and $T=\lambda^2\e^{-1}$,
\eqn
        \lefteqn{
        \Big(\Exp\Big[\|\phi_{n,N,\theta_{j-1}}(\theta_j)\|_2^{2\p}
        \Big]\Big)^{\frac1\p}
        }
        \label{aprioribd-2}
        \\
        &&\leq \;
        \frac{(c\lambda^2\e^{-1})^n}{\sqrt{n!}} \, + \,
        \Big((n!)\e^{\frac{1}{5}}+((n\p)!)^{\frac{1}{\p}}
        \e^{\frac{1}{5\p}}\Big) \,
        (\log\frac1\e)^{3} \,
        (c'\lambda^2\e^{-1}\log\frac1\e)^{n} \;,
        \nonumber
\eeqn
for finite constants $c$ and $c'=c'(T)$ which are independent of $L$.
\end{lemma}

This lemma is proved in Section {\ref{sect-lm4.2}}.

\begin{lemma}
\label{mainlm3}
For any fixed $\p\in2\N$ and
$T>0$, there exists a finite constant $c=c(T)$ independent of $L$ such that
\eqn
        \Big(\Exp\Big[ \|\widetilde\phi_{4N,N,\theta_{j-1}}(\theta_j)\|_2^{2\p}
        \Big]\Big)^{\frac1\p}\leq
        \frac{((4N\p)!)^{\frac{1}{2\p}}
        (\log\frac1\e)^{3}
        (c\lambda^2\e^{-1}\log\frac1\e)^{4N}}{\kappa^{2N}} \;.
\eeqn
\end{lemma}

The proof of this lemma is given in Section {\ref{sect-lm4.3}}.

We will make extensive use of the basic inequalities formulated in the following lemma.

\begin{lemma}
\label{lm:main-prop-est-1}
For $L$ sufficiently large (e.g. for (\ref{eq:L-bd-assump-1})),
\eqn
        \sup_{\alpha\in \Ie,k\in\Tor^3}\frac{1}{|\en(k)-\alpha-i\e|}
        &=&\frac1\e
        \label{prop-trivbd}
        \nonumber\\
        \int_{\Tor^3}\frac{dk}{|\en(k)-\alpha-i\e|}\; \; , \; \;
        \int_{\Ie}\frac{|d\alpha|}{|\en(k)-\alpha-i\e|}
        &<&c\log\frac1\e\;.
        \label{prop-logbd}
\eeqn
\end{lemma}

\prf
Clearly,
\eqn
        \int_{\LLs}\frac{dk}{|\en(k)-\alpha-i\e|} \; \leq \;
        \int_{\Tor^3}\frac{dk}{|\en(k)-\alpha-i\e|} \, + \, O\left(\frac{1}{\e^2|\LL|}\right) \;.
\eeqn
The bound $\int_{\Tor^3}\frac{dk}{|\en(k)-\alpha-i\e|}<c\log\frac1\e$ is proved in \cite{ch,erdyau}.
The remaining cases are evident.
\endprf

Moreover, we point out the following key property of the functions
$\phi_{n,N,\theta_{j-1}}(s)$.
From
\eqn
        \phi_{n,N,\theta}(s)&=&(-i\lambda)^{n-N}
        \int_{\R_+^{n-N+1}} ds_{0}\cdots ds_{n-N} \,
        \delta(\sum_{j=0}^{n-N}s_j-s)
        \nonumber\\
        &&\hspace{2cm}
        e^{is_0 \frac\Delta2}V_\omega\cdots V_\omega e^{is_{n-N}\frac\Delta2}
        V_\omega \phi_{N-1,\theta}\;,
        \label{phinNtheta-2}
\eeqn
where $s\in[\theta,\theta']$ and $\theta'-\theta=\frac t\kappa$, we find
\eqn
        (\widehat\phi_{n,N,\theta}(\theta'))(k_0)
        &=&\frac{(-\lambda)^{n-N} \, e^{(s-\theta)\kappa\e} }{2\pi i}
        \int_{\Ie}d\alpha \, e^{- i(s-\theta) \alpha  }
        \int_{(\LLs)^{n-N}} dk_{1}\cdots dk_{n-N}
        \nonumber\\
        &&
        \frac{1}{\en(k_0)-\alpha-i\kappa\e} \, \widehat V_\omega(k_1-k_0) \, \cdots
        \nonumber\\
        &&\hspace{1.5cm}\cdots \,
        \frac{1}{\en(k_{n-N})-\alpha-i\kappa\e} \,
        \widehat V_\omega(k_{n-N+1}-k_{n-N})
        \nonumber\\
        && \widehat\phi_{N-1,\theta}(k_{n-N+1}) \;,
\eeqn
recalling that $\e=\frac1t$, and with
\eqn
        \widehat\phi_{N-1,\theta}(k_{n-N+1})&=&
            \frac{(-\lambda)^N e^{\e\theta}}{2\pi i}
            \int_{\Ie}d\alpha \, e^{-i\alpha \theta}
        \int_{(\LLs)^N}\prod_{j=n-N+1}^{n}dk_j
        \nonumber\\
        && \frac{1}{\en(k_{n-N+1})-\alpha-  i\e}
        \, \widehat V_\omega(k_{n-N+2}-k_{n-N+1}) \, \cdots
        \nonumber\\
        &&\hspace{1cm}\cdots \, \widehat V_\omega(k_{n}-k_{n-1}) \,
        \frac{1}{\en(k_{n})-\alpha-  i\e} \,
        \widehat\phi_{0}(k_{n}) \;.
        \;\;\;\;\;\;
\eeqn
The key observation here is that there are $n-N+1$ propagators with
imaginary part $-i\kappa\e$ in the denominator, where $\kappa\e\gg\e$,
and $N$ propagators
where the corresponding imaginary part is $i\e$. Therefore,
we have a bound
\eqn
        \frac{1}{|\en(p)-\alpha-i\kappa\e|} \; \leq \; \frac{1}{\kappa\e}
        \; \ll \; \frac1\e\;
\eeqn
for $n-N+1$ propagators,
which is much smaller than the bound $\frac{1}{|\en(p)-\alpha-i\e|}\leq\frac1\e$.
This gain of a factor $\frac1\kappa$ as compared to (\ref{prop-trivbd})
is exploited in the time partitioning, and is applied systematically
in the proof of Lemma {\ref{mainlm3}}.

\section{Proof of Lemma {\ref{mainlm1}}}

We recall that
\eqn
        &&\Exptc \Big[\big| \langle \, \hJe \, , \, \widehat W_{t;n_1,n_2} \, \rangle  \big|^\p\Big]
        \nonumber\\
        &&\hspace{1cm}= \;
        \Exptc \Big[\Big(\big| \int_{ \LLsd \times \LLs}  \, dk \, dk' \,  \hJe(k-k',\frac{k+k'}{2}) \,
        \nonumber\\
        &&\hspace{6cm}
        \overline{\widehat\phi_{n_2,t}(k)} \,
        \widehat\phi_{n_1,t}(k') \big|^2\Big)^{\frac\p2}\Big] \;,
        \label{Econn-1}
\eeqn
and note that $\hJe$ forces $|k-k' \, ({\rm mod}\;2\Tor^3)|<c\lambda^2$, while
$|k+k'\, ({\rm mod}\;2\Tor^3)|$ is essentially unrestricted. Next, we introduce the
following multi-index notation. As $n_1,n_2$ will remain fixed in the
proof, let for brevity $n\equiv n_1$, and $\bar n\equiv n_1+n_2$.
For $j=1,\dots,r$, let \eqn
        \uk^{(j)}&:=&(k_0^{(j)},\dots,k_{\bar n+1}^{(j)})
        \nonumber\\
        d\uk^{(j)}&:=&\prod_{\ell=0}^{\bar n+1}dk_\ell^{(j)}
        \nonumber\\
        d\uk^{(j)}_{\hJe}&:=&\prod_{\ell=0}^{\bar n+1}dk_\ell^{(j)}
        \hJe(k^{(j)}_n-k^{(j)}_{n+1},\frac{k^{(j)}_n+k^{(j)}_{n+1}}{2})
        \nonumber\\
        K^{(j)}[\uk^{(j)},\alpha_j,\beta_j,\e ]&:=&
        \prod_{\ell=0}^n\frac{1}{\en(k_\ell^{(j)})-\alpha_j-i\e_j}
        \prod_{\ell'=n+1}^{\bar n+1}\frac{1}{\en(k_{\ell'}^{(j)})-\beta_j+i\e_j}
        \nonumber\\
        U^{(j)}[\uk^{(j)}]&:=&\prod_{\ell=1}^n
        \widehat V_\omega(k_\ell^{(j)}-k_{\ell-1}^{(j)})
        \prod_{\ell'=n+2}^{\bar n+1}
        \widehat V_\omega(k_{\ell'}^{(j)}-k_{\ell'-1}^{(j)})\;,
        \label{eq:Lr-graph-def-1}
\eeqn
where $\e_j:=(-1)^j\e$,
 $\alpha_j\in \Ie$, and $\beta_j\in\bar \Ie$ (the complex conjugate of $\Ie$).
On the last line, we note that
$\overline{\widehat V_\omega(k)}=\widehat V_\omega(-k)$.

Moreover, we introduce the notation
\eqn
        \ualpha \; := \; (\alpha_1,\dots,\alpha_\p) \; \; \; , \; \; \;
        d\ualpha \; := \; \prod_{j=1}^\p d\alpha_j \;,
\eeqn
and likewise for $\ubeta$, $\uxi$ and $d\ubeta$, $d\uxi$.

Then,
\eqn
        (\ref{Econn-1})&=&\frac{e^{2r\e t}\lambda^{r\bar n}}{(2\pi)^{2\p}}
        \int_{(\Ie\times \bar \Ie)^\p}
        d\ualpha \, d\ubeta \,e^{-it\sum_{j=1}^r(-1)^j(\alpha_j-\beta_j)}
        \nonumber\\
        && \hspace{1cm}
        \int_{(\Tor^3)^{(\bar n+2)\p}}
        \Big[ \, \prod_{j=1}^r d\uk^{(j)}_{\hJe} \Big] \,
        \Exptc \Big[\prod_{j=1}^r U^{(j)}[\uk^{(j)}]\Big]\,
        \nonumber\\
        && \hspace{2cm}
        \prod_{j=1}^r K^{(j)}[\uk^{(j)},\alpha_j,\beta_j,\e ] \,
        \widehat\phi_0^{(j)}(k_0^{(j)}) \, \overline{\widehat\phi_0^{(j)}(k_{\bar n+1}^{(j)})} \;,
        \label{End-1}
\eeqn
where
\eqn
        \widehat\phi_0^{(j)} \; := \; \left\{
        \begin{array}{rl}
        \widehat\phi_0&{\rm if}\;j\;{\rm is\;even}\\
        \overline{\widehat\phi}_0&{\rm if}\;j\;{\rm is \; odd}\;.
        \end{array}
        \right.
\eeqn
The expectation $\Exptc$
(defined in (\ref{Expnd-def-2}) below) in (\ref{End-1})
produces a sum of $O(( \bar n r )!)$
singular integrals
with complicated delta distribution insertions.
We organize them by use of {\em (Feynman) graphs}, which we define next, see also Figure 1.

We consider graphs comprising $r$ parallel, horizontal solid lines, which we refer to as {\em particle lines},
each containting $\bar n$ vertices enumerated from the left,
which account for copies of the random potential $\widehat V_\omega$.
Between the $n$-th and the $n+1$-th $\widehat V_\omega$-vertex, we a distinguished vertex is inserted
to account for the contraction with $\hJe$ (henceforth referred to as the
"$\hJe$-vertex"). Then, the $n$ edges on the left of the $\hJe$-vertex
correspond to the propagators in $\widehat\psi_{n,t}$ resp. $\overline{\widehat\psi_{n,t}}$, while the
$\bar n - n$ edges on the right correspond to those in
$\overline{\widehat\psi_{\bar n-n,t}}$ resp. $\widehat\psi_{\bar n-n,t}$. We shall refer to those edges,
labeled by the momentum variables $k_\ell^{(j)}$,
as {\em propagator lines}.

The expectation produces a sum over all possible
products of $\frac{r\bar n}{2}$ delta distributions,
each standing for one contraction between a pair of random potentials.
We connect every pair of mutually contracted random potentials
with a dashed {\em contraction line}. We then identify the contraction type with the
corresponding graph.

We remark that what is defined here as one particle line was referred to
as a pair of particle lines joined by a $\hJe$-, or respectively,
a $\delta$-vertex in \cite{ch,erdyau}. Thus, according to the terminology of
\cite{ch,erdyau}, we would here be discussing the case of $2r$
particle lines. Due to the different emphasis in the work at hand, the
convention introduced here appears to be more convenient.

We particularly distinguish the class of {\em completely disconnected
graphs}, in which random potentials are
mutually contracted only if they are located on the same particle line.
Clearly, all of its members possess $r$ connectivity components.

All other contraction types are referred to
as {\em non-disconnected graphs}.

A particular subfamily of non-disconnected graphs, referred to
as {\em 2-connected graphs}, is defined by the property that every connectivity
component has at least two particle lines.
Accordingly, we may now provide the following definitions which were in part already
anticipated in the preceding discussion.

\begin{definition}
\label{Expnd-def}
Let
\eqn
        \Expd\Big[\prod_{j=1}^r U^{(j)}[\uk^{(j)}]\Big]
        \; := \; \prod_{j=1}^r \Exp\Big[U^{(j)}[\uk^{(j)}]\Big]
\eeqn
include contractions among random potentials $\widehat V_\omega$
only if they lie on the same particle line. We refer to $\Expd$
as the expectation based on completely disconnected graphs.

We denote by
\eqn
        \Expnd\Big[\prod_{j=1}^r U^{(j)}[\uk^{(j)}]\Big] \; := \;
        \Exp\Big[\prod_{j=1}^r U^{(j)}[\uk^{(j)}]\Big]
        \, - \, \Expd\Big[\prod_{j=1}^r U^{(j)}[\uk^{(j)}]\Big]  \;,
        \label{Expnd-def-1}
\eeqn
the expectation based on non-disconnected graphs,
defined by the condition that there is at least one connectivity
component comprising more than one particle line.

Moreover, we refer to
\eqn
        \Exptc \Big[\prod_{j=1}^r U^{(j)}[\uk^{(j)}]\Big]
        \; := \; \Exp\Big[\prod_{j=1}^r \Big(U^{(j)}[\uk^{(j)}] \, - \,
        \Exp\Big[ U^{(j)}[\uk^{(j)}] \Big] \Big)\Big] \;
        \label{Expnd-def-2}
\eeqn
as the expectation based on 2-connected graphs.
\end{definition}

For $r\bar n\in 2\N$, let $\Pi_{r;\bar n,n}^{(\hJe)}$
denote the set of all graphs on $r\in\N$ particle lines, each containing
$\bar n$ $\widehat V_\omega$-vertices, and each
with the $\hJe$-vertex located between
the $n$-th and $n+1$-th $\widehat V_\omega$-vertex.
Then,
\eqn
        \Exp\Big[\prod_{j=1}^r U^{(j)}[\uk^{(j)}]\Big]
        \; = \; \sum_{\pi\in\Pi_{r;\bar n,n}^{(\hJe)}}
        \delta_\pi(\uk^{(1)},\dots,\uk^{(r)}) \;,
\eeqn
where $\delta_\pi(\uk^{(1)},\dots,\uk^{(r)})$ is defined as
follows. There are $\frac{r\bar n}{2}$ pairing contractions between
$\widehat V_\omega$-vertices in $\pi$. Every (dashed) contraction
line connects a random potential
$\widehat V_\omega(k_{n_1+1}^{(j_1)}-k_{n_1}^{(j_1)})$
with a random potential
$\widehat V_\omega(k_{n_2+1}^{(j_2)}-k_{n_2}^{(j_2)})$,
for some pair of multi-indices $((j_1,n_1),(j_2,n_2))$ determined by $\pi$, for which
\eqn
        \Exp\Big[\widehat V_\omega(k_{n_1+1}^{(j_1)}-k_{n_1}^{(j_1)})
        \widehat V_\omega(k_{n_2+1}^{(j_2)}-k_{n_2}^{(j_2)})\Big]
        \; = \; \delta(k_{n_1+1}^{(j_1)}-k_{n_1}^{(j_1)}
        +k_{n_2+1}^{(j_2)}-k_{n_2}^{(j_2)})\;.
\eeqn
Then,
$\delta_\pi(\uk^{(1)},\dots,\uk^{(\p)})$ is given by the
product of deltas (where $\delta(0)=|\LL|$ and $\delta(k)=0$ if $k\neq0$)
over all pairs of multi-indices $((n_1;j_1),(n_2;j_2))$
determined by $\pi$.

We refer to a graph with a single
connectivity component as a {\em completely connected graph}.
We denote the subset of $\Pi_{r;\bar n,n}^{(\hJe)}$
consisting of completely
connected graphs by $\Pi_{r;\bar n,n}^{(\hJe)\,conn}$.

Clearly, any graph $\pi\in\Pi_{r;\bar n,n}^{(\hJe)}$
is the disjoint union of its completely
connected components $\pi_j\in \Pi_{s_j;\bar n,n}^{(\hJe)\,conn}$
with $\sum s_j=\p$. Accordingly, (\ref{Econn-1}) factorizes into
the corresponding Feynman amplitudes, $\amp_{\hJe}(\pi)=\prod_j\amp_{\hJe}(\pi_j)$.

We may thus restrict our attention to completely connected graphs.

Let $\pi\in \Pi_{s;\bar n,n}^{(\hJe)\,conn}$ with $s\geq1$.
Its Feynman amplitude is given by
\eqn
        \widetilde{\amp}_{\hJe}(\pi)&:=&\frac{ \lambda^{s\bar n}
        e^{2s\e t} }{(2 \pi)^{2 s }}\int_{(\Ie\times \bar \Ie)^s}
        \, d\ualpha \, d\ubeta \, e^{-it\sum_{j=1}^s(-1)^j(\alpha_j-\beta_j)}
        \nonumber\\
        &&
        \int_{(\LLs)^{(\bar n+2)s}} \Big[ \, \prod_{j=1}^s d\uk^{(j)} \, \Big]
        \delta_\pi(\uk^{(1)},\dots,\uk^{(s)})\,
        \nonumber\\
        &&\int_{ (\Lambda_{L,\frac12}^*)^s} d\uxi \,
        \Big[ \, \prod_{j=1}^s\hJe(\xi_j,\frac{k_n^{(j)}+k_{n+1}^{(j)}}{2}) \,
        \delta(k_n^{(j)}-k_{n+1}^{(j)}-\xi_j) \, \Big]
        \nonumber\\
        && \hspace{1.5cm}
        \prod_{j=1}^s K^{(j)}[\uk^{(j)},\alpha_j,\beta_j,\e ]
        \, \widehat\phi_0^{(j)}(k_0^{(j)}) \,
        \overline{\widehat\phi_0^{(j)}(k_{\bar n+1}^{(j)})}
        \;.
\eeqn
Replacing $\int_{\LLs}dk$ by $\int_{\Tor^3}dk$, and the scaled Kronecker deltas on $\LLs$
by delta distributions on $\Tor^3$, we define
\eqn
        \amp_{\hJe}(\pi)&:=& \frac{ \lambda^{s\bar n} \,
        e^{2s\e t} }{(2\pi)^{2s}}\, \int_{(\Ie\times \bar \Ie)^s}
        \, d\ualpha \, d\ubeta \, e^{-it\sum_{j=1}^s(-1)^j(\alpha_j-\beta_j)}
        \nonumber\\
        &&
        \int_{(\Tor^3)^{(\bar n+2)s}} \, \Big[ \, \prod_{j=1}^s d\uk^{(j)} \, \Big]
        \delta_\pi(\uk^{(1)},\dots,\uk^{(s)})\,
        \nonumber\\
        && \int_{(2\Tor^3)^s} d\uxi \,
        \Big[ \, \prod_{j=1}^s\hJe(\xi_j,\frac{k_n^{(j)}+k_{n+1}^{(j)}}{2}) \,
        \delta(k_n^{(j)}-k_{n+1}^{(j)}-\xi_j) \, \Big]
        \nonumber\\
        && \hspace{1.5cm}
        \prod_{j=1}^s K^{(j)}[\uk^{(j)},\alpha_j,\beta_j,\e ] \,
        \widehat\phi_0^{(j)}(k_0^{(j)}) \,
        \overline{\widehat\phi_0^{(j)}(k_{\bar n+1}^{(j)})}
        \;,
        \label{Econn-2}
\eeqn
which is {\em independent} of $L$.
It is obvious that $\widetilde{\amp}_{\hJe}(\pi)$ is a discretization of $\amp_{\hJe}(\pi)$ on a
grid of lattice spacing $O(\frac1 L)$.
The discretization error is bounded in the following lemma.

\begin{lemma}
\eqn
        \Big| \, \widetilde{\amp}_{\hJe}(\pi) \, - \, \amp_{\hJe}(\pi) \, \Big|
        \; < \;  \frac{C(\bar n,\e)}{ |\LL|} \;,
\eeqn
where  $C(\bar n,\e)\leq O\left( \frac{(s \bar n)^{2}}{\e^{s(\bar n+2)+1}}\right)$.
\end{lemma}

\prf
The integrand in $\amp_{\hJe}(\pi)$
contains $s(\bar n+2)$ resolvent multipliers, each of which is bounded by $\frac1\e$ in $L^\infty(\Tor^3)$, and
by $\frac{1}{\e^2}$ in $C^1(\Tor^3)$.
It is demonstrated in our discussion below how to systematically integrate out all deltas in (\ref{Econn-2}).
Replacing the integral over $\Tor^3$ by the sum over $\LLs$ for
each momentum remaining  after integrating out the delta distributions in (\ref{Econn-2})
(using $\uxi$ to integrate out the functions $\hJe$, see (\ref{eq:hJe-int-bd-1})) yields
an error of order  $O(\frac{1}{|\LL|})$, multiplied with the sum of first derivatives of the integrand with respect to
each momentum. That integrand
is given by a product of $(\bar n + 2)s$ resolvent multipliers; differentiation with
respect to the momentum variables  yields a sum in which each term can be bounded by $\e^{-s(\bar n+2)-1}$.
Moreover,
this sum comprises no more than $(s(\bar n + 2 ))^2$ terms (where $s\leq\p$ is fixed).
\endprf

For the truncated Duhamel series (\ref{eq:Duham-exp-def-1}), we have to estimate
amplitudes of the form (\ref{Econn-2}) for $\bar n$  up to $\bar n\leq 4N \leq O(\log\frac1\e)$,
see (\ref{eq:Param-choice-1}) and (\ref{eq:Rn-4N-1}), (\ref{eq:Rn-4N-2})
(for $\bar n=4 N$, there are $2^\p$ propagators less, and the denominators
of some propagators have an imaginary part $\frac{1}{\kappa\e}$
instead of $\frac1\e$, see Section {\ref{sect-lm4.2}};
this only improves the bounds considered here). Thus, for
$L\gg \e^{-\p/\e}\geq\e^{-s/\e}$ (see (\ref{eq:L-bd-assump-1})),
the discretization error is smaller than $O(\e)$.
Accordingly, we shall henceforth only consider $\amp_{\hJe}(\pi)$, and assume $L$
to be sufficiently large for the discretization errors to be negligible in all cases
under consideration. In particular,
all bounds obtained in the sequel will be {\em uniform in} $L$,
and we recall that we are sending $L$ to $\infty$ first before taking any other limits.

The following key lemma is in part a joint result with Laszlo Erd\"os.

\begin{lemma}
\label{conngrbd-1}
Let $s\geq2$, $s\bar n\in 2\N$, and let
$\pi\in\Pi_{s;\bar n,n}^{(\hJe)\,conn}$ be a completely connected graph.
Then, there exists a finite constant $c=c(T)$ independent of $L$ such that
\eqn
        | \, \amp_{\hJe}(\pi) \, | \; \leq \;
        \e^{\frac{1}{5}} \, (\log\frac1\e)^{3} \, (c\lambda^2\e^{-1}
        \log\frac1\e)^{\frac{s\bar n}{2}} \;,
\eeqn
for every $T=\lambda^2\e^{-1}>0$.
\end{lemma}

\subsection{Classification of contractions}

For the proof of Lemma {\ref{conngrbd-1}}, we classify the
contractions among random potentials appearing in $\delta_\pi$ beyond the
typification introduced in \cite{erdyau} and \cite{ch}.

We define the following types of delta distributions.

\begin{definition}
A delta distribution of the form
\eqn
        \delta(k_{i+1}^{(j)}-k_i^{(j)}+k_{i'+1}^{(j)}-k_{i'}^{(j)})
        \; \; , \; \; |i-i'|\geq1
\eeqn
which connects the $i$-th with the $i'$-th vertex on the same particle line is called
an internal delta. The corresponding contraction line in the graph is
an internal contraction.

An internal delta with $|i-i'|=1$ is called an immediate recollision.

A delta distribution of the form
\eqn
        \delta(k_{i+1}^{(j)}-k_i^{(j)}+k_{i'+1}^{(j')}-k_{i'}^{(j')})
        \; \; , \; \; j\neq j'
\eeqn
which connects the $i$-th vertex on the $j$-th particle line
with the $i'$-th vertex on the $j'$-th particle line is called
a transfer delta. The corresponding contraction line is referred to as a
transfer contraction, and labeled by $((i;j),(i';j'))$. A vertex that
is adjacent to a transfer contraction is called a transfer vertex.

\end{definition}

\subsection{Reduction to the $L^4$-problem}

Assume that $s\geq2$.
Given a completely connected graph $\pi\in\Pi_{s;\bar n,n}^{(\hJe)\,conn}$,
we enumerate the transfer contraction lines by $\ell\in\{1,\dots,m\}$
(with $m$ denoting the number of transfer contraction lines in $\pi$).



We decompose $\pi\in\Pi_{s;\bar n,n}^{(\hJe)\,conn}$
into $s$ {\em reduced 1-particle lines} as follows, see also Figure 2.

Assume that the $\ell$-th transfer contraction is labeled by
$((i_\ell;j),(i'_\ell;j'))$. We replace the corresponding transfer
delta by the product
\eqn
    &&\delta(k_{i_\ell+1}^{(j)}-k_{i_\ell}^{(j)}+k_{i_\ell'+1}^{(j')}-k_{i_\ell'}^{(j')})
    \nonumber\\
    &&\hspace{3cm}\rightarrow \;
    \delta(k_{i_\ell+1}^{(j)}-k_{i_\ell}^{(j)}+u_\ell)
    \,
    \delta(k_{i_\ell'+1}^{(j')}-k_{i_\ell'}^{(j')}-u_\ell )
    \;,
\eeqn
where the first factor is attributed to
the  $j$-th, and the second factor to the $j'$-th particle line.
We say that $\delta(k_{i+1}^{(j)}-k_i^{(j)}+u_\ell) $ couples the
vertex $(i;j)$ to the new variable $u_\ell$.
We refer to $u_\ell$  as the {\em transfer momentum} corresponding to the
$\ell$-th transfer contraction line, and to $\delta(k_{i_\ell+1}^{(j)}-k_{i_\ell}^{(j)}+u_\ell)$
as the {\em reduced transfer delta} on the $j$-th particle line (parametrized by $u_\ell$).
We factorize every transfer delta, and associate each reduced transfer delta to the corresponding
contraction line.




Let $\uu^{(j)}$ comprise all transfer momenta $u_\ell$ which couple
to a transfer vertex on the $j$-th particle line. We define
\eqn
    \delta_{int}(\uk^{(j)}) \; := \; \prod_{{\rm internal \; deltas}}
    \delta(k_{i+1}^{(j)}-k_{i}^{(j)}+k_{i'+1}^{(j)}-k_{i'}^{(j)})
\eeqn
and
\eqn
    \delta^{(j)} (\uu^{(j)},\uk^{(j)}) \; :=  \; \delta_{int}(\uk^{(j)})
    \prod_{u_\ell\;{\rm belonging \; to \;}\uu^{(j)}
    \atop  {\rm on \;} j-{\rm th \; particle \; line } }\delta(k_{i_\ell+1}^{(j)}-k_{i_\ell}^{(j)}\pm u_\ell) \;,
\eeqn
which comprises all deltas on the $j$-th particle line, including the corresponding factors
from the modified transfer deltas.

Moreover, every vertex carries a factor $\lambda$.

\begin{definition}
The $j$-th reduced 1-particle graph $\pi_j(\uu^{(j)})$ comprises the
$j$-th particle line, $\bar n$ $V_\omega$-vertices, one $\hJe$-vertex,
all internal contractions, but none of the transfer contraction lines. The
transfer vertices carry the reduced transfer deltas,
and are parametrized by $\uu^{(j)}$.

Accordingly, we refer to
\eqn
        \amp_{\hJe}(\pi_j(\uu^{(j)}))& :=& \frac{ \lambda^{\bar n}
        e^{2s\e t} }{(2\pi )^{2}}\int_{\Ie\times \bar \Ie }
        d\alpha_j d\beta_j \,e^{-it (-1)^j(\alpha_j-\beta_j)}
        \nonumber\\
        &&
        \int_{(\Tor^3)^{\bar n+2}}  d\uk^{(j)}
        \delta^{(j)} (\uu^{(j)},\uk^{(j)})
        \nonumber\\
        && \int_{2\Tor^3} d\xi_j\,
        \hJe(\xi_j ,\frac{ k_n^{(j)} + k_{n+1}^{(j)} }{2})
        \,\delta( k_n^{(j)} -  k_{n+1}^{(j)} -\xi_j)
        \nonumber\\
        &&\hspace{1cm}
        K^{(j)}[ \uk^{(j)} ,\alpha_j,\beta_j,\e ]
        \,
        \widehat\phi_0^{(j)}(k_0^{(j)}) \,
        \overline{\widehat\phi_0^{(j)}( k_{\bar n+1}^{(j)} )}
    \label{eq:ampj-def-1}
\eeqn
as the $j$-th reduced 1-particle amplitude.
\end{definition}


The amplitude $\amp_{\hJe}(\pi)$ is obtained from the product of all reduced
1-particle amplitudes, by integrating over the transfer momenta.

\begin{lemma}
\label{lm:factor-1}
(Factorization lemma)
Assume that $\pi\in\Pi_{s;\bar n,n}^{(\hJe)\,conn}$, for $s\geq2$, carries
the transfer momenta $\uu=(u_1,\dots,u_m)$. Let $\pi_j(\uu^{(j)})$, for $j=1,\dots,s$, denote the
$j$-th reduced 1-particle graph. Then,
\eqn
    \amp_{\hJe}(\pi) \; = \; \int du_1\dots du_m \prod_{\ell=1}^s
    \amp_{\hJe}(\pi_j(\uu^{(j)}))  \;.
\eeqn
Notably, every $u_\ell$ in $\uu$ appears in precisely two different reduced 1-particle
amplitudes (once with each sign).
\end{lemma}

Next, we reduce the problem for $s\geq2$ to the problem $s=2$ (corresponding to a
completely connected $L^4$-graph).


To this end, let us assume that $\pi$ contains $m$ transfer contractions, carrying the transfer
momenta $\uu=(u_1,\dots,u_m)$. Then, by (\ref{lm:factor-1}),
\eqn
        |\amp_{\hJe}(\pi)| \; \leq \; \int d\uu \,
        \Big[\prod_{j=1}^s|\amp_{\hJe}(\pi_j(\uu^{(j)}))|\Big] \;,
\eeqn
where $\uu^{(j)}$ denotes the subset of $m_j$ transfer momenta which
couple to the $j$-th particle line.
Moreover, let $\uu^{(j;i)}$ denote the subset of transfer momenta in $\uu^{(j)}$
belonging to transfer contractions between the $j$-th and the $i$-th reduced 1-particle line.
We recall that every transfer momentum appears in precisely two reduced 1-particle amplitudes.
Hence,
\eqn
        \uu^{(j;i)} \; = \; \uu^{(i;j)} \; \; {\rm for \; all} \; i\ne j \; ,
        \; \; {\rm and} \; \; \uu^{(i;i)} \; = \; \emptyset \; \;
        {\rm  for \; all} \; i \;.
\eeqn
Assuming that $\uu^{(s-1;s)}\neq\emptyset$ (possibly after relabeling
the particle lines),
\eqn
        |\amp_{\hJe}(\pi)|&\leq&\int d\uu^{(1;2)}\cdots d\uu^{(1;s)} d\uu^{(2;3)}\cdots
        \cdots d\uu^{(s-2;s-1)} d\uu^{(s-2;s)} d\uu^{(s-1;s)}
        \nonumber\\
        &&\hspace{1cm}
        \Big[\prod_{j=1}^s|\amp_{\hJe}(\pi_j(\uu^{(j;1)},\dots,\uu^{(j;j-1)},\uu^{(j;j+1)},\dots,
        \uu^{(j;s)}))|\Big]
        \nonumber\\
        &\leq&\Big[\int d\uu^{(1;2)}\cdots d\uu^{(1;s)}
        |\amp_{\hJe}(\pi_1(\uu^{(1;2)},\dots,\uu^{(1;s)}))|\Big]
        \nonumber\\
        &&\sup_{\uu^{(1;2)}}\Big[\int d\uu^{(2;3)}\cdots d\uu^{(2;s)}
        |\amp_{\hJe}(\pi_2(\uu^{(2;1)},\uu^{(2;3)},\dots,\uu^{(2;s)}))|\Big]
        \nonumber\\
        &&\hspace{4cm}\cdots\cdots
        \nonumber\\
        &&\sup_{{\uu^{(\ell;s-1)},\uu^{(\ell;s)} \atop 1\leq \ell\leq s-2}}
        \Big[\int d\uu^{(s-1;s)}
        |\amp_{\hJe}(\pi_{s-1}(\uu^{(s-1;1)},\dots,\uu^{(s-1;s)}))|
        \nonumber\\
        &&\hspace{4cm}
        |\amp_{\hJe}(\pi_{s}(\uu^{(s;1)},\dots,\uu^{(s;s-1)}))|\Big]
        \nonumber\\
        &=&\Big[\prod_{j=1}^{s-2} A_j \Big] \, B \;,
        \label{eq:ampj-bd-aux-2}
\eeqn
where
\eqn
        A_j&:=& \sup_{ {\uu^{(i;j)} \atop 1\leq i< j}}\Big[\int d\uu^{(j;j+1)}\cdots d\uu^{(j;s)}
        |\amp_{\hJe}(\pi_j(\uu^{(j;1)},\dots,\uu^{(j;j-1)},
        \nonumber\\
        &&\hspace{6cm}\uu^{(j;j+1)},\dots,\uu^{(j;s)}))|\Big]
\eeqn
for $1\leq j\leq s-1$, and
\eqn
        A_s \; := \; \sup_{ {\uu^{(i;s)} \atop 1\leq i< s}}
        |\amp_{\hJe}(\pi_s(\uu^{(s;1)},\dots \dots,\uu^{(s;s-1)}))|
\eeqn
($A_{s-1}$ and $A_s$ are used in the a priori bound of Lemma {\ref{lm:apriori-1}} below).
Moreover,
\eqn
        B&:=&\sup_{{\uu^{(\ell;s-1)},\uu^{(\ell;s)} \atop 1\leq \ell\leq s-2}}
        \Big[\int d\uu^{(s-1;s)}
        |\amp_{\hJe}(\pi_{s-1}(\uu^{(s-1;1)},\dots,\uu^{(s-1;s)}))|
        \nonumber\\
        &&\hspace{4cm}
        |\amp_{\hJe}(\pi_{s}(\uu^{(s;1)},\dots,\uu^{(s;s-1)}))|\Big]
\eeqn
corresponds to a completely connected $L^4$-graph.
We note that
\eqn
        B \; \leq \; A_{s-1} \, A_s
\eeqn
is evident.

Next, we estimate the terms $A_j$.

\begin{lemma}
\label{lm:trunc-est-1}
Assume that the $j$-th truncated particle line contains $m_j$ transfer
deltas, carrying the transfer momenta $\uu^{(j)}$.
Let $\uu^{(j)}=(u_1^{(j)},\dots,u_{m_j}^{(j)})$, according to an arbitrary enumeration of
the transfer vertices.

Let $a\in\N_0$ and  $0\leq a\leq m_j$, and arbitrarily partition
$\uu^{(j)}$ into
$\uu_1^{(j)}$ and $\uu_\infty^{(j)}$, where $\uu_1^{(j)}$ contains $a$, and $\uu_\infty^{(j)}$ contains $m_j-a$ transfer momenta.
Then,
\eqn
        \sup_{\uu_\infty^{(j)}}\int d\uu_1^{(j)}
        |\amp_{\hJe}(\pi_j(\uu^{(j)}))| \; < \;
        (c \, \lambda)^{\bar n}\e^{-\frac{\bar n+m_j}{2}+a}(\log\frac1\e)^{\frac{\bar n-m_j}{2}+a+2} \;,
\eeqn
for a constant $c$ which is independent of $\e$.
\end{lemma}

\prf
For notational convenience, we may, without any loss of generality, assume that
\eqn
        \uu_1^{(j)} \; := \; (u_1^{(j)},\dots,u_{a}^{(j)})
        \; \; , \; \;
        \uu_\infty^{(j)} \; := \; (u_{a+1}^{(j)},\dots,u_{m_j}^{(j)})
\eeqn
(by possibly relabeling the transfer momenta in $\uu^{(j)}$).
We recall the definition of $\amp(\pi_j(\uu^{(j)}))$ from (\ref{eq:ampj-def-1}),
and note that $\pi_j(\uu^{(j)})$ contains $m_j$ vertices carrying reduced transfer deltas,
$\bar n - m_j\in 2\N$ vertices that are adjacent to an internal contraction line, and one $\hJe$-vertex.
Clearly,
\eqn
        \lefteqn{
        \int d\uu^{(j)}_1|\amp_{\hJe}(\pi_j(\uu^{(j)}))|
        \; \leq \;
        \frac{ \lambda^{\bar n} } {(2\pi)^2} \,
        e^{2 \e t}  \, \Big( \, \int_{2\Tor^3}d\xi \, \sup_v|\hJe(\xi,v) |   \, \Big)
        }
        \nonumber\\
        &&\hspace{1cm}\sup_\xi \int_{\Ie\times \bar \Ie }
        |d\alpha_j| \, |d\beta_j|
        \int_{(\Tor^3)^{\bar n+2+a}}  d\uu^{(j)}_1 d\uk^{(j)}
        \delta^{(j)} (\uu^{(j)},\uk^{(j)})
        \,\delta( k_n^{(j)} -  k_{n+1}^{(j)} -\xi )
        \nonumber\\
        &&\hspace{3cm}
        |K^{(j)}[ \uk^{(j)} ,\alpha_j,\beta_j,\e ]|
        \,
        |\widehat\phi_0^{(j)}(k_0^{(j)})| \,
        |\widehat\phi_0^{(j)}( k_{\bar n+1}^{(j)} )|  \;,
    \label{eq:ampj-bd-aux-1}
\eeqn
and we recall (\ref{eq:hJe-int-bd-1}).
Adding the arguments of all delta distributions, we find the momentum conservation condition
\eqn
        k_{\bar n+1}^{(j)} \; = \; k_0^{(j)} \, + \, \xi \, + \, \sum_{i=1}^{m_j} (\pm u_i^{(j)}) \;,
        \label{eq:red1part-ends-1}
\eeqn
linking the momenta at both ends of the reduced 1-particle graph.
We replace the delta belonging to the vertex $(\bar n;j)$ by
$\delta(k_{\bar n+1}^{(j)} - k_0^{(j)} - \xi - \sum_{i=1}^{m_j} (\pm u_i^{(j)}))$,
irrespective of it being an internal or a reduced transfer delta, and
remove it from $\delta^{(j)} (\uu^{(j)},\uk^{(j)})$.
We integrate out the $\hJe$-delta
$\delta( k_n^{(j)} -  k_{n+1}^{(j)} -\xi )$  using the variable $k_{n+1}^{(j)}$, and the delta
$\delta(k_{\bar n+1}^{(j)} - k_0^{(j)} - \xi - \sum_{i=1}^{m_j} (\pm u_i^{(j)}))$
using the variable $k_{\bar n+1}^{(j)}$.
It follows that if $1\leq n < \bar n$,
\eqn
        |(\ref{eq:ampj-bd-aux-1})|&\leq& C  \lambda^{\bar n}  \sup_\xi  \,
        \int_{(\Tor^3)^{ a}}  d\uu^{(j)}_1 \, \int dk_0^{(j)} \,|\widehat\phi_0^{(j)}(k_0^{(j)})| \,
        |\widehat\phi_0^{(j)}( k_0^{(j)}+ \xi+\sum_{i=1}^{m_j} (\pm u_i^{(j)}) )|
        \nonumber\\
        &&
        \int_{\Ie\times \bar \Ie }
        |d\alpha_j| \, |d\beta_j| \, F[\uu^{(j)},k_0^{(j)},\xi,\alpha_j,\beta_j,\e]
        \nonumber\\
        && \hspace{1.5cm}
        \frac{ 1 }
        {|\en(k_0^{(j)})-\alpha_j-i\e| \, |\en(k_0^{(j)}+ \xi+\sum_{i=1}^{m_j} (\pm u_i^{(j)}))-\beta_j+i\e|}
        \nonumber
\eeqn
where
\eqn
        F[\uu^{(j)},k_0^{(j)},\xi,\alpha_j,\beta_j,\e]
        & :=&
        \int_{(\Tor^3)^{\bar n -1}}   dk_n^{(j)} \, d\widetilde\uk^{(j)}
        \widetilde \delta^{(j)}(\uu^{(j)},\widehat\uk^{(j)},\xi)\,
        \nonumber\\
        && \hspace{0.5cm}
        \frac{|\widetilde K^{(j)}[ \widetilde\uk^{(j)} ,\alpha_j,\beta_j,\e ]|}
    {|\en(k_n^{(j)})-\alpha_j-i\e| \, |\en(k_n^{(j)}+\xi)-\beta_j+i\e|} \; , \; \; \; \;\;\;
\eeqn
with
\eqn
    \widehat\uk^{(j)} \; := \; (k_0^{(j)},k_n^{(j)},\widetilde\uk^{(j)})
    \; \; \; , \; \; \;
        \widetilde\uk^{(j)} \; := \; (k_1^{(j)},\dots,k_{n-1}^{(j)},k_{n+2}^{(j)},\dots,k_{\bar n}^{(j)})
\eeqn
and
\eqn
        \widetilde K^{(j)}[\widetilde\uk^{(j)},\alpha_j,\beta_j,\e ]&:=&
        \prod_{\ell=1}^{n-1}\frac{1}{\en(k_\ell^{(j)})-\alpha_j-i\e_j}
        \nonumber\\
        &&\hspace{1cm}
        \prod_{\ell'=n+2}^{\bar n }\frac{1}{\en(k_{\ell'}^{(j)})-\beta_j+i\e_j} \;.
\eeqn
Here, $\widetilde \delta^{(j)}(\uu^{(j)},\widehat\uk^{(j)},\xi)$
is obtained from $\delta^{(j)}(\uu^{(j)},\uk^{(j)})$ by omitting
the delta distribution belonging to the vertex $(\bar n, j)$, and by substituting
$k_{n+1}^{(j)}\rightarrow k_{n}^{(j)}-\xi$.
%
Splitting
\eqn
            |\widehat\phi_0^{(j)}(k_0^{(j)})| \,
        |\widehat\phi_0^{(j)}( k_0^{(j)}+ u )| \; \leq \; \frac12|\widehat\phi_0^{(j)}(k_0^{(j)})|^2
            +\frac12|\widehat\phi_0^{(j)}( k_0^{(j)}+ u )|^2 \;,
\eeqn
we find
\eqn
        |(\ref{eq:ampj-bd-aux-1})| \; \leq \; (I)+(II) \;,
\eeqn
where
\eqn
        (I)&\leq& C \, \lambda^{\bar n}   \,\Big[ \int dk_0^{(j)} |\widehat\phi_0^{(j)}(k_0^{(j)})|^2  \Big] \,
        \Big[\sup_{k_0^{(j)}}\int_{\Ie }
        |d\alpha_j| \frac{1}{|\en(k_0^{(j)})-\alpha_j-i\e|}\Big]\,
        \nonumber\\
        &&
        \sup_\xi\sup_{\alpha_j}\sup_{k_0^{(j)}}
        \int_{\bar\Ie}|d\beta_j|\, \int_{(\Tor^3)^{ a}}  d\uu^{(j)}_1
        \,
        \frac{F[\uu^{(j)},k_0^{(j)},\xi,\alpha_j,\beta_j,\e]}
        {|\en(k_0^{(j)}+ \xi+\sum_{i=1}^{m_j} (\pm u_i^{(j)}))-\beta_j+i\e|}
        \nonumber\\
        &\leq&C \, \lambda^{\bar n} \,  \|\widehat\phi_0\|_2^2 \,  \log\frac1\e \,
        \label{eq:red1p-I-aux-est-1}
            \\
            &&
        \sup_\xi \sup_{\alpha_j}\sup_{k_0^{(j)}}
        \int_{\bar\Ie}|d\beta_j|\, \int_{(\Tor^3)^{ a}}  d\uu^{(j)}_1
        \,
        \frac{F[\uu^{(j)},k_0^{(j)},\xi,\alpha_j,\beta_j,\e]}
        {|\en(k_0^{(j)}+ \xi+\sum_{i=1}^{m_j} (\pm u_i^{(j)}))-\beta_j+i\e|}  \;.
        \nonumber
\eeqn
and
\eqn
        (II)&\leq& C \, \lambda^{\bar n} \,  \sup_\xi  \,
        \int_{(\Tor^3)^{ a}}  d\uu^{(j)}_1 \, \int dk_0^{(j)} \,
        |\widehat\phi_0^{(j)}( k_0^{(j)}+ \xi+\sum_{i=1}^{m_j} (\pm u_i^{(j)}) )|^2
        \nonumber\\
        &&
        \int_{\Ie\times \bar \Ie }
        |d\alpha_j| \, |d\beta_j|\frac{1}{|\en(k_0^{(j)})-\alpha_j-i\e|}\,
        \nonumber\\
        &&\hspace{3cm}
        \frac{F[\uu^{(j)},k_0^{(j)},\xi,\alpha_j,\beta_j,\e]}
        {|\en(k_0^{(j)}+ \xi+\sum_{i=1}^{m_j} (\pm u_i^{(j)}))-\beta_j+i\e|}
        \nonumber\\
        &\leq&C \,  \lambda^{\bar n}   \,\Big[ \int dk_0^{(j)} |\widehat\phi_0^{(j)}(k_0^{(j)})|^2  \Big]\,
        \Big[\sup_{k_0^{(j)}}\int_{\bar\Ie }
        |d\beta_j| \frac{1}{|\en(k_0^{(j)})-\beta_j-i\e|}\Big]\,
        \nonumber\\
        &&
        \sup_\xi\sup_{\beta_j}\sup_{k_0^{(j)}}
        \int_{ \Ie}|d\alpha_j|\, \int_{(\Tor^3)^{ a}}  d\uu^{(j)}_1
            \, \frac{F'[\uu^{(j)},k_0^{(j)},\xi,\alpha_j,\beta_j,\e]}
        {|\en(k_0^{(j)}- \xi-\sum_{i=1}^{m_j} (\pm u_i^{(j)}))-\alpha_j+i\e|}
        \nonumber\\
        &\leq& C \,  \lambda^{\bar n} \, \|\widehat\phi_0\|_2^2 \,  \log\frac1\e
        \label{eq:red1p-II-aux-est-1}\\
            &&
        \sup_\xi\sup_{\beta_j}\sup_{k_0^{(j)}}\int_{ \Ie}|d\alpha_j|\,
        \int_{(\Tor^3)^{ a}}  d\uu^{(j)}_1 \,
        \frac{F'[\uu^{(j)},k_0^{(j)},\xi,\alpha_j,\beta_j,\e]}
        {|\en(k_0^{(j)}- \xi-\sum_{i=1}^{m_j} (\pm u_i^{(j)}))-\alpha_j+i\e|} \;.
        \nonumber
\eeqn
We have here applied a shift
$k_0^{(j)}\rightarrow k_0^{(j)}- \xi-\sum_{i=1}^{m_j} (\pm u_i^{(j)})$
which induces $F\rightarrow F'$ in the obvious way.
We note that this only affects the delta distributions belonging to
the vertices $(1,j)$ and $(\bar n,j)$ in
$\widetilde\delta^{(j)}(\uu^{(j)},\widehat\uk^{(j)},\xi)$ of $F$.

We focus on $(I)$, the case of $(II)$ is analogous.
We have
\eqn
        \lefteqn{
        \sup_\xi\sup_{\alpha_j}\sup_{k_0^{(j)}}\int_{ \Ie}|d\beta_j|\,
        \int_{(\Tor^3)^{ a}}  d\uu^{(j)}_1 \frac{F[\uu^{(j)},k_0^{(j)},\xi,\alpha_j,\beta_j,\e] }
            {|\en(k_0^{(j)}+ \xi+\sum_{i=1}^{m_j} (\pm u_i^{(j)}))-\beta_j+i\e|}
        }
       \\
        &=&
        \sup_\xi\sup_{\alpha_j}\sup_{k_0^{(j)}}
        \int_{ \Ie}|d\beta_j|\,\int \, d\uu^{(j)}_1 \, dk_n^{(j)} \, d\widetilde \uk^{(j)} \,
        \frac{
        |\widetilde K^{(j)}[ \widetilde\uk^{(j)} ,\alpha_j,\beta_j,\e ]|  }
            {|\en(k_0^{(j)}+ \xi+\sum_{i=1}^{m_j} (\pm u_i^{(j)}))-\beta_j+i\e|}
        \nonumber\\
        &&\hspace{3cm}\frac{\widetilde \delta^{(j)}(\uu^{(j)},\widehat\uk^{(j)},\xi)}
    {|\en(k_n^{(j)})-\alpha_j-i\e| \, |\en(k_n^{(j)}+\xi)-\beta_j+i\e|}\;.
        \nonumber\\
        &\leq&
        \Big[\sup_{k_n^{(j)}}\int_{ \Ie}|d\beta_j| \frac{1}{|\en(k_n^{(j)}+\xi)-\beta_j+i\e|}\Big]
        \nonumber\\
        &&
        \sup_\xi\sup_{\alpha_j}\sup_{k_0^{(j)}}
        \,\int \, d\uu^{(j)}_1 \, dk_n^{(j)} \, d\widetilde \uk^{(j)} \,
        \frac{
        |\widetilde K^{(j)}[ \widetilde\uk^{(j)} ,\alpha_j,\beta_j,\e ]|  }
            {|\en(k_0^{(j)}+ \xi+\sum_{i=1}^{m_j} (\pm u_i^{(j)}))-\beta_j+i\e|}
        \nonumber\\
        &&\hspace{5cm}\frac{\widetilde \delta^{(j)}(\uu^{(j)},\widehat\uk^{(j)},\xi)}
    {|\en(k_n^{(j)})-\alpha_j-i\e| \, }\;.
\eeqn
Next, we integrate out the reduced transfer deltas:

\begin{itemize}
\item
\underline{The case $1\leq \ell\leq a$:} If  $i_\ell<\bar n$, we integrate out the corresponding transfer deltas
$\delta(k_{i_\ell+1}^{(j)}-k_{i_\ell}^{(j)}\pm u_\ell^{(j)})$
using the transfer momenta $u_\ell^{(j)}$ (the components of $\uu^{(j)}_1$).
Then, for each such $\ell$, we use the variable $k_{i_\ell+1}^{(j)}$
(on the right of the corresponding transfer vertex, according to our conventions)
to estimate the corresponding propagator in $L^1$,
\eqn
    \int  \frac{dk_{i_\ell+1}^{(j)}}{|e(k_{i_\ell+1}^{(j)})-\gamma \pm i\e|} \; < \; c \, \log\frac1\e
\eeqn
(where $\gamma$ denotes $\alpha_j$ or $\beta_j$). If the $\bar n$-th vertex is a transfer
vertex, and $i_\ell=\bar n$,
we recall that the corresponding transfer delta
has already been integrated out using the momentum  $k_{\bar n+1}^{(j)}$, and
replaced by the delta enforcing (\ref{eq:red1part-ends-1}).
Accordingly, we use $u_\ell$ for the estimate
\eqn
    \int \frac{du_\ell}{|\en(k_0^{(j)}+ \xi+\sum_{i=1}^{m_j} (\pm u_i^{(j)}))-\beta_j+i\e|}
    \; \leq \; c \, \log\frac1\e \;,
\eeqn
noting that the propagator in the integrand (supported on the edge initially labeled by $k_{\bar n+1}^{(j)}$)
is the only one depending on $u_\ell$.
Thus, in this step, $a$ propagators are in total estimated in $L^1$ by $c\log\frac1\e$, irrespectively
whether there is $\ell\leq a$ with $i_\ell=\bar n$ or not.

\item
\underline{The case $a<\ell\leq m_j$:} If $i_\ell<\bar n$, we integrate out the corresponding reduced transfer deltas
$\delta(k_{i_\ell+1}^{(j)}-k_{i_\ell}^{(j)}\pm u_\ell^{(j)})$ using
the variable $k_{i_\ell+1}^{(j)}$ on the right of the associated vertex $(i_\ell;j)$.
We then estimate each of the corresponding propagators by
\eqn
    \sup_{k_{i_\ell+1}^{(j)}}\frac{1}{|e(k_{i_\ell+1}^{(j)})-\gamma \pm i\e|} \; \leq \; \frac 1\e
\eeqn
in $L^\infty$.
If $i_\ell=\bar n$, we again note that the corresponding transfer delta
has already been integrated out using the momentum  $k_{\bar n+1}^{(j)}$.
For the propagator supported on the edge labeled by $k_{\bar n+1}^{(j)}$, we use
\eqn
    \frac{1}{|\en(k_0^{(j)}+ \xi+\sum_{i=1}^{m_j} (\pm u_i^{(j)}))-\beta_j+i\e|}
    \; \leq \;  \frac1\e \; .
    \label{eq:barn-res-est-Linft-1}
\eeqn
Thus, in this step, $m_j-a$ propagators are in total estimated in $L^\infty$ by $\frac1\e$, irrespectively of
whether there is $\ell> a$ with $i_\ell=\bar n$ or not.

\end{itemize}

We summarize that out of the $\bar n+2$ momenta in $\uk^{(j)}$, we have used $k_0^{(j)}$,
$k_{n+1}^{(j)}$, and $k_{\bar n+1}^{(j)}$ to begin with. Moreover, if the
$\bar n$-th vertex is a transfer vertex, we have used another $m_j-1$ components of $\uk^{(j)}$
to either integrate out transfer deltas, or to estimate propagators in $L^1$.
On the other hand, if the
$\bar n$-th vertex is an internal vertex, we have, to this end, used $m_j$ components of $\uk^{(j)}$.

We also note that out of the $\bar n+2$ propagators, $a$ have been estimated by $\frac1\e$ in
$L^\infty$, and $m_j-a+2$ (two from the integrals in $\alpha_j$ and $\beta_j$)  by $c\log\frac1\e$
in $L^1$.

Next, we introduce a spanning tree $T$ on $\pi_j(\uu^{(j)})$, which
contains all internal contraction lines, but none of the transfer vertices, and none of the
$m_j-a+2$ edges carrying propagators that were already estimated above in $L^1$ or $L^\infty$.
Thus, in particular, $T$ does not contain the propagator edges
corresponding to the momenta $k_0^{(j)}$,  $k_{n+1}^{(j)}$ and $k_{\bar n+1}^{(j)}$.
We then call $T$ {\em admissible}.

Thus, we distinguish the following cases:
\begin{itemize}
\item
\underline{The $\bar n$-th vertex is an internal vertex:} The corresponding internal
delta has already been replaced by the delta enforcing (\ref{eq:red1part-ends-1}), and integrated
out using $k_{\bar n+1}^{(j)}$.
Accordingly, we use the estimate
 (\ref{eq:barn-res-est-Linft-1})  for the propagator on its right.
Out of the remaining $\bar n-2-m_j$ momenta in $\uk^{(j)}$,
we use $\frac{\bar n-m_j}{2}-1$ momenta supported on $T$ to
integrate out the remaining internal deltas,
and we estimate the corresponding propagators in $L^\infty$ by $\frac1\e$.
There remain $\frac{\bar n-m_j}{2}-1$ momenta for $L^1$-bounds
on the corresponding propagators.

\item
\underline{The $\bar n$-th vertex is a transfer vertex:}
Out of the remaining $\bar n-3-m_j$ momenta in $\uk^{(j)}$, we use $\frac{\bar n-m_j}{2}$ momenta supported on $T$ to
integrate out the internal deltas, and we estimate the corresponding propagators in $L^\infty$ by $\frac1\e$.
There remain $\frac{\bar n-m_j}{2}-2$ momenta for $L^1$-bounds
on the corresponding propagators.
\end{itemize}

With
the r\^oles of the propagators on
the edges labeled by $k_0^{(j)}$ and $k_{\bar n+1}^{(j)}$ interchanged,
the discussion for the term $(II)$ is fully analogous to the one of $(I)$.

Summarizing, $\frac{\bar n -m_j}{2}+(m_j-a)=\frac{\bar n+m_j}{2}-a$ propagators are in total bounded in $L^\infty$,
and $\frac{\bar n -m_j}{2}+2 +a$ in $L^1$.
In conclusion, we obtain
\eqn
    \sup_{\uu_\infty^{(j)}}\int  d\uu_1^{(j)}|\amp_{\hJe}(\pi_j(\uu^{(j)}))|
    \; < \; (c\lambda)^{\bar n}\e^{-\frac{\bar n+m_j}{2}+a}(\log\frac1\e)^{\frac{\bar n-m_j}{2}+a+2} \;,
    \label{eq:intuu-amp-j-bd-main-1}
\eeqn
as claimed.
The cases $n=0$ and $n=\bar n$ are similar, and also yield (\ref{eq:intuu-amp-j-bd-main-1}).
This can be proved with minor modifications of the arguments explained above, and will not
be reiterated.
This concludes the proof.
\endprf

\begin{lemma}
\label{lm:apriori-1}
Let $\pi\in\Pi_{s;\bar n,n}^{(\hJe)\,conn}$.
We then have the a priori bound
\eqn
        |\amp_{\hJe}(\pi)| \; < \; (\log\frac1\e)^{3s}
        (c\lambda^2\e^{-1}\log\frac1\e)^{\frac{s\bar n}{2}} \; .
\eeqn
%
\end{lemma}

\prf
From (\ref{eq:ampj-bd-aux-2}), we have
\eqn
        |\amp_{\hJe}(\pi)|&\leq&\prod_{j=1}^s A_j \;.
        \label{eq:ampj-bd-aux-3}
\eeqn
Using (\ref{eq:ampj-bd-aux-3}) and Lemma {\ref{lm:trunc-est-1}}, we get
\eqn
        |\amp_{\hJe}(\pi)|&\leq&(c\lambda)^{s \bar n}\prod_{j=1}^s
        \e^{-\frac{\bar n+m_j}{2}+a_j}(\log\frac1\e)^{\frac{\bar n-m_j}{2}+a_j+2}
        \nonumber\\
        &<&(c\lambda)^{s\bar n}
        \e^{-\frac{s\bar n+\sum_j m_j}{2}+\sum_j a_j}(\log\frac1\e)^{\frac{s\bar n-\sum_j m_j}{2}+\sum_j a_j +2s} \;.
\eeqn
We observe that $m=\sum_{j}m_j\in2\N_0$ is twice the number of transfer contractions in $\pi$,
since it counts the number of transfer vertices.
Moreover, $\sum_j a_j=\frac{m}{2}$ because to every transfer contraction, we associate one resolvent
estimated in $L^1$ and one estimated in $L^\infty$, and $\sum_j a_j$ counts
those estimated in $L^1$. This implies the asserted bound.
%
\endprf

Next, we estimate the term $B$ in  (\ref{eq:ampj-bd-aux-2}), and show that exploiting
the connectedness of a pair of particle lines, one gains a factor $\e^{\frac15}$ over the
bound $B\leq A_{s-1}A_s$ inferred from Lemma {\ref{lm:trunc-est-1}}.



\begin{lemma}
Assume that the reduced 1-particle lines $\pi_j(\uu^{(j)})$ and $\pi_{j'}(\uu^{(j')})$
have $m_{j;j'}$ common transfer momenta $\uu^{(j;j')}$. Let $\widetilde \uu^{(i;j)}$
denote the $m_j+m_{j'}-2m_{j;j'}$
transfer momenta appearing in either $\uu^{(j)}$ or $\uu^{(j')}$, but not in both.
Moreover, assume that $\widehat \phi_0$ satisfies the  concentration of
singularity condition  (\ref{eq:phi0-sing-conc-1}).
Then,
\eqn
        &&\sup_{\widetilde\uu^{(j;j')} }\int d\uu^{(j;j')}|\amp_{\hJe}(\pi_j(\uu^{(j)}))|
        \,|\amp_{\hJe}(\pi_{j'}(\uu^{(j')}))| \;
        \nonumber\\
        &&\hspace{2cm}\leq \;\lambda^{2\bar n}\e^{\frac{1}{5}-\bar n-\frac{m_j+m_{j'}}{2}+m_{j;j'}}
        (c \log\frac1\e)^{ \bar n -\frac{m_j+m_{j'}}{2}+m_{j;j'}+4}
        \label{eq:L4-bd-main-1}
\eeqn
which improves the corresponding a priori bound by a factor $\e^{\frac{1}{5}}$.
\end{lemma}

\prf
To estimate the l.h.s. of (\ref{eq:L4-bd-main-1}), we use $L^\infty-L^1$-bounds in the variables
$\uu^{(j;j')}$, with the exception of one transfer momentum, which we denote by $u$.
Thereby, we cut all but one transfer lines between the
$j$-th and the $j'$-th reduced 1-particle line.

One straightforwardly obtains (\ref{eq:L4-bd-main-1}) if it is possible to identify
a subgraph in the expression for (\ref{eq:L4-bd-main-1}) that corresponds to the "crossing integral"
\eqn
        &&\sup_{\gamma_i\in \Ie}\sup_{k\in\Tor^3}
        \Big[\int dp_1 dp_2 \,\frac{1}{|\en(p )-\gamma_1-i\e_1| \,|\en(q)-\gamma_2-i\e_1| }
        \label{eq:crossing-1}\\
        &&\hspace{3.5cm}
        \frac{1}{\,|\en(p-q+k)-\gamma_3-i\e | }\Big]
        \; \leq \; c\,\e^{-\frac45}(\log\frac1\e)^3 \;,
        \nonumber
\eeqn
see Lemma 3.11 in \cite{ch}. Here, one of the three resolvents would have been estimated in
$L^\infty$ by $\frac1\e$ in the a priori bound. There is a gain of a factor $\e^{\frac15}$
because the singularities which contribute most to (\ref{eq:crossing-1}) are concentrated
in tubular $\e$-neighborhoods of level surfaces of $\en$,
whose intersections are of small measure (the curvature of the level surfaces
of the energy function $\en$ plays a crucial role for this result).

On each reduced 1-particle line, we identify the contraction structure based on internal
deltas, see also  Figure 3. As explained in detail in \cite{erdyau,ch}, the only possible cases are
(we are here omitting the labels $j$, $j'$ of the reduced 1-particle lines):
\begin{itemize}
\item
The internal contractions of the reduced 1-particle line define a ladder graph decorated with
progressions of immediate recollisions. That
is, every internal contraction is either an immediate recollision
(a contraction between neighboring internal vertices, possibly with transfer vertices
located inbetween), or a rung of the ladder contracting a vertex labeled by $i\leq n$ with
a vertex labeled by $i'>n$. For any pair of rung contractions labeled by $(i_1,i'_1)$ and $(i_2,i'_2)$,
one has $i_1<i_2$, and $i'_1>i'_2$ (no crossing of rungs).
These were denoted "simple graphs" in \cite{erdyau,ch}.
\item
Otherwise, one can identify at least one nesting or crossing subgraph. A
pair of internal deltas $\delta(k_{i_1+1}-k_{i_1}+k_{i_1'+1}-k_{i_1'})$
and $\delta(k_{i_2+1}-k_{i_2}+k_{i_2'+1}-k_{i_2'})$ defines a {\em nesting subgraph}
if $i_1<i_2<i_2'<i_1'$, and either $i_1'\leq n$ or $i_1>n$.
It defines a {\em crossing subgraph} if
$i_1<i_2<i_1'<i_2'$, and either $i_2'\leq n$ or $i_1>n$.
\end{itemize}

In (\ref{eq:L4-bd-main-1}), one can identify a crossing subintegral of the form (\ref{eq:crossing-1})
in the following situations:
\begin{itemize}
\item
One of the reduced 1-particle graphs contains a nesting or crossing subgraph consisting of
internal contraction lines, similarly as in \cite{erdyau,ch}. Then, one can completely disconnect
the $j$-th and the $j'$-th reduced 1-particle line by $L^\infty-L^1$-estimates in
$\uu^{(j;j')}$, and one still gains a factor $\e^{\frac15}$ from (\ref{eq:crossing-1}).
\item
Both reduced 1-particle subgraphs correspond to ladder graphs with immediate
recollision insertions (denoted "simple graphs" in \cite{erdyau,ch}), but there is at least one
transfer contraction between the $j$-th and the $j'$-th reduced 1-particle line
whose ends are located either between rungs of the ladder (that is, not on the left or
right of the outermost rung contraction $\delta(k_{i_*+1}-k_{i_*}+k_{i_*'+1}-k_{i_*'})$, where
$i_*$ is the smallest, and $i_*'$ is the largest index appearing in any rung contraction on
the given reduced 1-particle line) and/or inside an
immediate recollision subgraph.
The integral over the associated transfer momentum $u$ then
produces a subintegral of the form  (\ref{eq:crossing-1}),
and one gains a factor $\e^{\frac15}$.
\end{itemize}

The crossing estimate cannot be applied in its basic form  (\ref{eq:crossing-1}) when
the $j$-th and the $j'$-th reduced 1-particle graphs have ladder structure,
and every transfer contraction between the $j$-th and the $j'$-th reduced
1-particle graph is adjacent to at least one vertex on the left or right of the
outermost rung contraction,
which is also not located inside an immediate recollision subgraph.
Then, the corresponding integrals do not only involve propagators, but
also $\widehat\phi_0$, which itself typically exhibits singularities.

The situation is most difficult to handle if
both of the adjacent transfer vertices are of that type. Then, the only
subintegral with crossing structure has the form
\eqn
        A_\e&:=&\int d\alpha_1 \, d\alpha_2\int  dp\,  dq\, du\,
        |\widehat\phi_0(p )|\,|\widehat\phi_0(p+u+k)|\,|\widehat\phi_0(q)|\,|\widehat\phi_0(q-u)|\,
        \nonumber\\
        &&\hspace{1cm}\frac{1}{|\en(p)-\alpha_1-i\e |\,|\en(p+u+k)-\beta_1+i\e |}
        \nonumber\\
        &&\hspace{2cm}
        \frac{1}{
        \,|\en(q)-\alpha_2-i\e |\,|\en(q-u)-\beta_2+i\e |} \;.
        \label{eq:Ae-def-1}
\eeqn
This expression is obtained from partitioning the integrals on the r.h.s. of
(\ref{eq:L4-bd-main-1}) in the same way as in the proof of Lemma {\ref{lm:trunc-est-1}}
(we recall that on each reduced 1-particle line,
one of the energy parameters $\alpha_j$ or $\beta_j$ is always used to
estimate a propagator neighboring to either $\widehat\phi_0$ or $\overline{\widehat\phi}_0$ in $L^1$).
In (\ref{eq:Ae-def-1}), singularities of $|\widehat\phi_0|$
may overlap with those of the neighboring resolvents; crossing structures then also depend
on the singularity structure of $|\widehat\phi_0|$.
If we argue as in the proof of Lemma {\ref{lm:trunc-est-1}}, we would use two momentum integrals for
$\|\widehat\phi_0\|_{L^2(\Tor^3)}^4$,
and the remaining integrals ($\alpha_1$, $\alpha_2$, and the third momentum)
to bound three resolvents in $L^1(\Tor^3)$ by $c\log\frac1\e$ so that one resolvent is
estimated in $L^\infty(\Tor^3)$ by $\frac1\e$.
Thereby, one gets
\eqn
        A_\e<c\e^{-1}(\log\frac1\e)^3\|\widehat\phi_0\|_{L^2(\Tor^3)}^4 \;.
        \label{eq:Ae-apriori-bd-1}
\eeqn
The remaining terms contributing to the l.h.s. of (\ref{eq:crossing-1}) are estimated in the same way
as in the proof of Lemma {\ref{lm:trunc-est-1}} (i.e. by introduction of a spanning
tree, and use of $L^1-L^\infty$-bounds on the propagators),
whereby one again arrives at the expression for the a
priori bound, which is the r.h.s. of  (\ref{eq:crossing-1}) without the $\e^{\frac15}$-factor. We shall not
repeat the detailed argument.

To prove (\ref{eq:L4-bd-main-1}), we improve (\ref{eq:Ae-apriori-bd-1}) by
\eqn
    A_\e \; \leq  \; c(T)\, \e^{-\frac45} \, (\log\frac1\e)^4   \;,
\eeqn
where the constant depends only on the macroscopic time $T>0$.
Our proof uses the $\eta$-concentration property of the WKB initial data $\widehat\phi_0$.
We do not know if for general $L^2$ initial data, or for WKB initial conditions without
any restrictions on the phase function $S$, (\ref{eq:Ae-apriori-bd-1}) can be improved.

We recall the concentration of singularity condition
(\ref{eq:phi-0-dec-1}) - (\ref{eq:phi0-sing-conc-1}), by which
\eqn
    \widehat\phi_0(k)\; = \; f_\infty(k) \, + \, f_{sing}(k) \;,
\eeqn
where
\eqn
    \|f_\infty\|_\infty \; < \; c   \;,
\eeqn
and
\eqn
    \| \, |f_{crit}|*|f_{crit}|\, \|_{L^2(\Tor^3)}
    \; = \; \| \, | f_{sing} |^\vee \,\|_{\ell^4(\Z^3)}^2 \; \leq \; c' \, \eta^{\frac45}
    \label{eq:conc-sing-2}
\eeqn
for  constants $c$, $c'$ that are uniform in $\eta$.

We observe that $A_\e$ has the form
\eqn
    A_\e[g_1,g_2,g_3,g_4]&=&\int_{\Ie^2} d\alpha_1\,d\alpha_2\,
    \big\langle \widehat g_1*\widehat g_2 \, , \, \widehat g_3*\widehat g_4\big\rangle_{L^2(\Tor)}
    \nonumber\\
    &=&\int_{\Ie^2} d\alpha_1\,d\alpha_2\,
    \big\langle g_1 \, g_2 \, , \, g_3 \, g_4\big\rangle_{\ell^2(\Z^3)}
    \nonumber\\
    &=&\sum_{r_i\in\{\infty,crit\}}\, A_\e[g_{1,r_1},g_{2,r_2},g_{3,r_3},g_{4,r_4}] \;,
    \label{eq:Aeps-def-1}
\eeqn
where
$\widehat g_1(p):=\frac{1}{|\en(p)-\alpha_1-i\e|}|\widehat\phi_0(p)|$, etc., and where
$\widehat g_{i,r}$ is obtained from replacing $\widehat\phi_0$ by $f_r$ in $\widehat g_i$, for
$r\in\{\infty,crit\}$.
The corresponding terms can then be bounded as follows.

First of all, if $r_i=\infty$ for $i=1,\dots,4$,
\eqn
        \lefteqn{
        A_\e[g_{1,\infty},g_{2,\infty},g_{3,\infty},g_{4,\infty}]
        }
        \nonumber\\
        &\leq&\|f_\infty\|_\infty^4  \sup_{\alpha_j,\beta_j}\int dp\,  dq\, du\,
        \frac{1}{|\en(p)-\alpha_1-i\e |\,|\en(p+u)-\beta_1+i\e |}
        \nonumber\\
        &&\hspace{2cm}
        \frac{1}{
        \,|\en(q)-\alpha_2-i\e |\,|\en(q-u)-\beta_2+i\e |}
        \nonumber\\
        &<&c \, \e^{-\frac45}(\log\frac1\e)^4\;,
\eeqn
using (\ref{eq:crossing-1}).

If $r_i=crit$ for one value of $i$,
\eqn
        \lefteqn{
        A_\e[g_{1,crit},g_{2,\infty},g_{3,\infty},g_{4,\infty}]
        }
        \nonumber\\
        &\leq&\|f_\infty\|_\infty^3 \,
        \Big[\int dp \, |f_{crit}(p)| \, \Big]\,
        \nonumber\\
        &&\hspace{2cm}
        \Big[ \sup_{p} \frac{1}{\en(p+u)-\beta_1+i\e} \Big] \,
        \Big[\int  dq\,\frac{1}{ |\en(q)-\beta_2+i\e |}\Big]
        \nonumber\\
        &&\hspace{2cm}
        \Big[\sup_p\int_{\Ie}d\alpha_1\frac{1}{|\en(p)-\alpha_1-i\e |}\Big]\,
        \Big[\sup_q\int_{\Ie}d\alpha_2\frac{1}{|\en(q)-\alpha_2-i\e |}\Big]\,
        \nonumber\\
        &<&c \, \e^{-1} \, \eta^{\frac25} \, (\log\frac1\e)^4\;,
\eeqn
using $\|f_{crit}\|_{L^1(\Tor^3)} \, \leq\, c \, \eta^{\frac25}$,
which follows from
\eqn
        \|f_{crit}\|_{L^1(\Tor^3)}^2
        &=&\int dp \, du \, |f_{crit}(p)|\,|f_{crit}(u)|
        \nonumber\\
        &=&\int dp \, du \, |f_{crit}(p)|\,|f_{crit}(u-p)|
        \nonumber\\
        &\leq&\Big(\int du \Big[\int dp\,  |f_{crit}(p)|\,|f_{crit}(u-p)| \, \Big]^2\Big)^{\frac12}
        \Big(\int du\Big)^{\frac12}
        \nonumber\\
        &=&\|\,|f_{crit}|*|f_{crit}|\,\|_{L^2(\Tor^3)}
        \nonumber\\
        &\leq&c\eta^{\frac{4}{5}} \;,
        \label{eq:L1-fcrit-est-1}
\eeqn
see  (\ref{eq:conc-sing-2}), and where we have used ${\rm Vol}(\Tor^3)=1$.
The remaining cases $r_1=r_3=r_4=\infty$, $r_2=crit$,
etc., are similar.

If $r_i=crit$ for two values of $i$,
\eqn
        \lefteqn{
        \int_{\Ie}d\alpha_1\,d\alpha_2\,
        A_\e[g_{1,crit},g_{2,crit},g_{3,\infty},g_{4,\infty}]
        }
        \nonumber\\
        &\leq&\|f_\infty\|_\infty^2 \int_{\Ie^2} d\alpha_1\,d\alpha_2\,\int dp\,  dq\, du\,
        |f_{crit}(p)|\,|f_{crit}(p+u)|
        \frac{1}{|\en(p)-\alpha_1-i\e |\,}
        \nonumber\\
        &&\hspace{2cm}
        \frac{1}{|\en(p+u)-\beta_1+i\e |
        \,|\en(q)-\alpha_2-i\e |\,|\en(q-u)-\beta_2+i\e |}
        \nonumber\\
        &\leq&\|f_\infty\|_\infty^2 \,
        \Big[\int dp \, |f_{crit}(p)|\Big]\,
        \Big[\int du \, |f_{crit}(p+u)| \Big] \, \e^{-1}
        \nonumber\\
        &&\hspace{2cm}
        \Big[\sup_u\int dq
        \frac{1}{|\en(q-u)-\beta_2+i\e |}\Big]
        \nonumber\\
        &&\hspace{2cm}
        \Big[\sup_p\int_{\Ie}d\alpha_1\frac{1}{|\en(p)-\alpha_1-i\e |}\Big]\,
        \Big[\sup_q\int_{\Ie}d\alpha_2\frac{1}{|\en(q)-\alpha_1-i\e |}\Big]\,
        \nonumber\\
        &<&c\,\e^{-1} \, \eta^{\frac45}  \, (\log\frac1\e)^3\;,
\eeqn
again using (\ref{eq:L1-fcrit-est-1}).
The cases
$r_1=r_3 =\infty$, $r_2=r_4=crit$, etc., are similar.

If $r_i=crit$ for three values of $i$,
\eqn
        \lefteqn{
        \int_{\Ie}d\alpha_1\,d\alpha_2\,
        A_\e[g_{1,crit},g_{2,crit},g_{3,crit},g_{4,\infty}]
        }
        \nonumber\\
        &\leq&\|f_\infty\|_\infty \int_{\Ie^2} d\alpha_1\,d\alpha_2\,\int dp\,  dq\, du\,
        |f_{crit}(p)|\,|f_{crit}(p+u)|\,|f_{crit}(q)|
        \nonumber\\
        &&\hspace{2cm}
        \frac{1}{|\en(p)-\alpha_1-i\e |\,|\en(p+u)-\beta_1+i\e |}
        \nonumber\\
        &&\hspace{4cm}
        \frac{1}{\,|\en(q)-\alpha_2-i\e |\,|\en(q-u)-\beta_2+i\e |}
        \nonumber\\
        &\leq&\|f_\infty\|_\infty \, \|f_{crit}\|_{L^1(\Tor^3)}^3 \, \e^{-2} \,
        \nonumber\\
        &&\hspace{2cm}
        \Big[\sup_p\int_{\Ie}d\alpha_1\frac{1}{|\en(p)-\alpha_1-i\e |}\Big]\,
        \Big[\sup_q\int_{\Ie}d\alpha_2\frac{1}{|\en(q)-\alpha_2-i\e |}\Big]\,
        \nonumber\\
        &\leq&c \, \e^{-2} \, \eta^{\frac{6}{5}} \, (\log\frac1\e)^2 \,
\eeqn
using (\ref{eq:L1-fcrit-est-1}).
The remaining cases are similar.

Finally, if $r_i=crit$ for all values of $i$,
\eqn
        \lefteqn{
        \int_{\Ie}d\alpha_1\,d\alpha_2\,
        A_\e[g_{1,crit},g_{2,crit},g_{3,crit},g_{4,crit}]
        }
        \nonumber\\
        &\leq& \int_{\Ie^2} d\alpha_1\,d\alpha_2\,\int dp\,  dq\, du\,
        |f_{crit}(p)|\,|f_{crit}(p+u)|\,|f_{crit}(q)|\,|f_{crit}(q-u)|
        \nonumber\\
        &&\hspace{2cm}
        \frac{1}{|\en(p)-\alpha_1-i\e |\,|\en(p+u)-\beta_1+i\e |}
        \nonumber\\
        &&\hspace{4cm}
        \frac{1}{\,|\en(q)-\alpha_2-i\e |\,|\en(q-u)-\beta_2+i\e |}
        \nonumber\\
        &\leq& \e^{-2} \,\| \, |f_{crit}|*|f_{crit}| \, \|_{L^2(\Tor^3)}^2
        \nonumber\\
        &&\hspace{2cm}
        \Big[\sup_p\int_{\Ie}d\alpha_1\frac{1}{|\en(p)-\alpha_1-i\e |}\Big]\,
        \Big[\sup_q\int_{\Ie}d\alpha_2\frac{1}{|\en(q)-\alpha_1-i\e |}\Big]\,
        \nonumber\\
        &<&c \, \e^{-2} \, \eta^{\frac85} \, (\log\frac1\e)^2\;,
\eeqn
using (\ref{eq:conc-sing-2}).

We recall that $\eta=\lambda^2=T\e$, where $\e=\frac1t$ is the inverse microscopic time,
and $T=\lambda^2t$ is the macroscopic time.

Collecting the estimates on (\ref{eq:Aeps-def-1}) derived above, we find that for any $T>0$,
there is a constant $c(T)<\frac{c}{T^2}$ such that
\eqn
    A_\e \; \leq \; c(T) \, \e^{-\frac45}(\log\frac1\e)^4 \;.
\eeqn
This estimate improves (\ref{eq:Ae-apriori-bd-1}) by a factor $\e^{\frac15}$, as claimed,
and establishes (\ref{eq:L4-bd-main-1}).
\endprf

Using the arguments used in the proof of Lemma {\ref{lm:apriori-1}},
one hereby also establishes Lemma {\ref{conngrbd-1}}.

Moreover, we find the following bounds.

\begin{lemma}
\label{Nondiscsum-est1}
Let $r\in2\N$, and let $\Pi_{r;\bar n,n}^{(\hJe)\, 2-conn},
\Pi_{r;\bar n,n}^{(\hJe)\, n-d}\subset
\Pi_{r;\bar n,n}^{(\hJe)}$ denote the subclasses
of 2-connected and non-disconnected
graphs, respectively. Then,
for every $T=\lambda^2\e^{-1}>0$, there exists a finite constant $c=c(T)$ such that
\eqn
        \sum_{\pi\in\Pi_{r;\bar n,n}^{(\hJe)\, 2-conn}}|\amp_{\hJe}(\pi)|
        \; \leq \;
        (r\bar n)! \, \e^{\frac{1}{5}}(\log\frac1\e)^{3r } (c\lambda^2\e^{-1}
        \log\frac1\e)^{\frac{r\bar n}{2}} \;.
\eeqn
and
\eqn
        \sum_{\pi\in\Pi_{r;\bar n,n}^{(\hJe)\, n-d}}|\amp_{\hJe}(\pi)|
        \; \leq \;
        (r\bar n)!\,\e^{\frac{1}{5}}(\log\frac1\e)^{3r } (c\lambda^2\e^{-1}
        \log\frac1\e)^{\frac{r\bar n}{2}} \;.
\eeqn
\end{lemma}

\prf
This follows immediately from Lemma {\ref{conngrbd-1}} and
the fact that $\Pi_{r;\bar n,n}^{(\hJe)}$ contains no more than $(r\bar n)! 2^{r\bar n}$ graphs.
\endprf

\begin{lemma}
For any fixed $\p\geq2$, $\p\in2\N$,  $n\leq N$,
and $T>0$, there exists a finite constant $c=c(T)$ such that
\eqn
        \Big(\Exp\Big[ \|\phi_{n,t}\|_2^{2\p}
        \Big]\Big)^{\frac1\p} \; \leq \;
        ((2n\p)!)^{\frac{1}{\p}}(\log\frac1\e)^{3}
        (c\lambda^2\e^{-1}\log\frac1\e)^{n} \;.
        \label{Expphin-L2}
\eeqn
\end{lemma}

\prf
This is proved in the same way as the a priori bound of Lemma {\ref{lm:apriori-1}}.
The only modification is that the $\hJe$-delta is replaced by $\delta(k_n^{(j)}-k_{n+1}^{(j)})$
on every particle line. We note that the expansion for (\ref{Expphin-L2}) contains disconnected graphs.
\endprf

\section{Proof of Lemma {\ref{mainlm2}}}
\label{sect-lm4.2}

Based on the previous discussion, is straightforward to see that
\eqn
        \lefteqn{
        \Exp\big[\|\phi_{n,N,\theta_{m-1}}(\theta_m)\|_2^{2r}\big]
        }
        \nonumber\\
        &&
        \; = \; \frac{e^{2r ( 1 +  \theta_{m-1}\e)}}{(2\pi)^{2r}}
        \int_{(\Ie\times \bar \Ie)^r}
        \prod_{j=1}^r \, d\alpha_j \, d\beta_j \, e^{-i\theta_m\sum_{j=1}^r(-1)^j(\alpha_j-\beta_j)}
        \nonumber\\
        &&\hspace{1cm} \,
        \int_{(\Tor^3)^{(\bar n+2)r}}\Big[ \, \prod_{j=1}^r d\uk^{(j)}
        \delta(k_n^{(j)}-k_{n+1}^{(j)}) \, \Big] \, \lambda^{r\bar n} \,
        \Exp\Big[\prod_{j=1}^r U^{(j)}[\uk^{(j)}]\Big] \,
        \nonumber\\
        &&\hspace{2cm}
        \,
        \Big[\, \prod_{j=1}^r K^{(j)}_{n,N,\kappa}
        [\uk^{(j)},\alpha_j,\beta_j,\e ] \,\Big] \,
        \widehat\phi_0^{(j)}(k_0^{(j)}) \,
        \overline{\widehat\phi_0^{(j)}(k_{\bar n+1}^{(j)})} \;.
        \label{ExpphinN-1}
\eeqn
(using $(\theta_m-\theta_{m-1})\kappa \e=1$) where
\eqn
        K^{(j)}_{n,N,\kappa}
        \;:=\;\frac{1}{(\en(k_{n}^{(j)})-\alpha_j-i\e_j)
        (\en(k_{n+1}^{(j)})-\beta_j+i\e_j)} \,
        \widetilde K^{(j)}_{n,N,\kappa} \;,
\eeqn
and
\eqn
        &&\widetilde K^{(j)}_{n,N,\kappa}
        [\uk^{(j)},\alpha_j,\beta_j,\e ]
        \nonumber\\
        &&\hspace{1cm}:= \;
        \prod_{\ell_1=0}^{n-N}
        \frac{1}{\en(k_{\ell_1}^{(j)})-\alpha_j-i\kappa\e_j}
        \prod_{\ell_2=n-N+1}^{n-1}
        \frac{1}{\en(k_{\ell_2}^{(j)})-\alpha_j-i\e_j}
        \label{Kj-def-1}\\
        &&\hspace{1.5cm} \,
        \prod_{\ell_3=n+2}^{n+N}
        \frac{1}{\en(k_{\ell_3}^{(j)})-\beta_j+i\e_j}
        \prod_{\ell_4=n+N+1}^{2n+1}
        \frac{1}{\en(k_{\ell_4}^{(j)})-\beta_j+i\kappa\e_j} \;,
        \nonumber
\eeqn
see also (\ref{eq:Lr-graph-def-1}) and the discussion following (\ref{phinNtheta-2}).
We refer to $\delta(k_n^{(j)}-k_{n+1}^{(j)})$, which replaces the $\hJe$-delta, as
the "{\em $L^2$-delta}", since it is responsible for the $L^2$-inner
product on the left hand side of (\ref{ExpphinN-1}).
The expression (\ref{ExpphinN-1}) can be estimated in the same way as the integrals
(\ref{End-1}) considered above,  however,
we are now considering the full instead of the non-disconnected expectation.

As before, $\Exp\big[\prod_{j=1}^r U^{(j)}[\uk^{(j)}]\big]$ in (\ref{ExpphinN-1}) decomposes into a sum of
products of delta distributions, which we represent by Feynman graphs.
By the notational conventions introduced after (\ref{End-1}), we
have $\bar n=2n$.
We let $\Pi_{r;\bar n,n}$ denote the set of graphs on $r$ particle lines,
each containing $\bar n$ vertices from copies of
the random potential $\widehat V_\omega$, and with the $L^2$-delta located
between the $n$-th and the $n+1$-st
$\widehat V_\omega$-vertex. For $\pi\in\Pi_{r;n,\bar n}$, let $\amp_\delta(\pi)$
denote the amplitude corresponding to the graph $\pi$, given by the
integral obtained from replacing
$\Exp\big[\prod_{j=1}^r U^{(j)}[\uk^{(j)}]\big]$ in (\ref{ExpphinN-1}) by
$\delta_\pi(\uk^{(1)},\dots,\uk^{(r)})$ (the product of delta distributions
corresponding to the contraction graph $\pi$). The subscript in $\amp_\delta$
implies that instead of $\hJe$ as before, we now have the $L^2$-delta
at the distinguished vertex.

Let $\Pi_{r;\bar n,n}^{conn}$ denote the subclass of $\Pi_{r;\bar n,n}$ of
completely connected graphs.

\begin{lemma}
\label{conngrbd-2}
Let $s\geq2$, $s\in\N$,   and let
$\pi\in\Pi_{s;2n,n}^{conn}$ (that is, $\bar n=2n$) be a completely connected graph.
Then, for every $T=\lambda^2\e^{-1}>0$, there exists a finite constant $c=c(T)$ such that
\eqn
        |\amp_\delta(\pi)| \; \leq \; \e^{\frac{1}{5}}(\log\frac1\e)^{3s }(c\lambda^2\e^{-1}
        \log\frac1\e)^{ sn} \;.
\eeqn
\end{lemma}

\prf
The proof is completely analogous to the one given for Lemma {\ref{conngrbd-1}}
(using $\frac{1}{\kappa\e}\leq\frac1\e$),
and will not be reiterated here.
\endprf

In contrast to the situation in Lemma {\ref{conngrbd-1}},
the expectation in (\ref{ExpphinN-1}) contains completely disconnected
graphs, which satisfy
\eqn
        \sum_{\pi\in\Pi_{r;\bar n,n}^{disc}}|\amp_\delta(\pi)|
        \; \leq \; \Big(\sum_{\pi\in\Pi_{1;\bar n,n}^{conn}}|\amp_\delta(\pi)|\Big)^r\;.
\eeqn
We invoke the following bound from \cite{ch}
(the continuum version is proved in \cite{erdyau}).

\begin{lemma}
Let $\bar n=2n$.
Then, for a constant $c$ independent of $T$,
\eqn
        \sum_{\pi\in\Pi_{1;\bar n,n}^{conn}}|\amp_\delta(\pi)| \; \leq \;
        \frac{(c\lambda^2\e^{-1})^n}{\sqrt{n!}}
        \, + \, (n!) \, \e^{\frac{1}{5}} \, (\log\frac1\e)^{3} \, (c\lambda^2\e^{-1}
        \log\frac1\e)^{n} \;.
        \label{ch-mainbd-1}
\eeqn
\end{lemma}

The term $\frac{(c\lambda^2\e^{-1})^n}{\sqrt{n!}}$ bounds the
contribution from decorated ladder diagrams, while the term that
carries an additional $\e^{\frac{1}{5}}$-factor
is obtained from   crossing and nesting type subgraphs.
The proof of (\ref{ch-mainbd-1}) is presented in
detail in \cite{ch} and \cite{erdyau}.
The number of non-ladder graphs is bounded by $n!\,2^n$, hence the
factor $n!$.

The sum over non-disconnected graphs can be estimated by the same bound
as in Lemma {\ref{Nondiscsum-est1}}. The result is formulated in the
following lemma.

\begin{lemma}
\label{Nondiscsum-est2}
Let $r\in2\N$, $\bar n=2n$, and $\Pi_{r;\bar n,n}^{n-d}\subset
\Pi_{r;\bar n,n}$ denote the subclass
of non-disconnected
graphs. Then, for every $T>0$,
there exists a finite constant $c=c(T)$ such that
\eqn
        \sum_{\pi\in\Pi_{r;\bar n,n}^{n-d}}|\amp_{\delta}(\pi)| \; \leq \;
        (r\bar n)! \, \e^{\frac{1}{5}} \, (\log\frac1\e)^{3r } \, (c\lambda^2\e^{-1}
        \log\frac1\e)^{\frac{r\bar n}{2}} \;.
\eeqn
\end{lemma}

Combining (\ref{ch-mainbd-1}) with Lemma {\ref{Nondiscsum-est2}},
and applying the Minkowski inequality,
the statement of Lemma {\ref{mainlm2}} follows straightforwardly.

\section{Proof of Lemma {\ref{mainlm3}}}
\label{sect-lm4.3}

For $r\in2\N$, $s\in[\theta_{m-1},\theta_m]$, $\theta_m-\theta_{m-1}=\frac{t}{\kappa}$,
$n=4N$, and $\bar n=8N$, one gets
\eqn
        \lefteqn{
        \Exp\big[\|\widetilde\phi_{4N,N,\theta_{m-1}}(s)\|_2^{2r}\big]
        }
        \nonumber\\
        &&
        \; = \;  \frac{e^{2r((s-\theta_m)\kappa+\theta_m)\e }}{(2\pi)^{2r}}
        \int_{(\Ie\times \bar \Ie)^r}
        \prod_{j=1}^r d\alpha_jd\beta_j \,e^{-is\sum_{j=1}^r(-1)^j(\alpha_j-\beta_j)}
        \nonumber\\
        &&
        \hspace{1cm} \,
        \int_{(\Tor^3)^{(\bar n+2)r}}
        \Big[ \, \prod_{j=1}^r d\uk^{(j)}
        \delta(k_{4N}^{(j)}-k_{4N+1}^{(j)}) \, \Big] \, \lambda^{r\bar n} \,
        \Exp\Big[\prod_{j=1}^r U^{(j)}[\uk^{(j)}]\Big]\,
        \nonumber\\
        &&
        \hspace{2cm}
        \,
        \Big[ \, \prod_{j=1}^r \widetilde K^{(j)}_{4N,N,\kappa}
        [\uk^{(j)},\alpha_j,\beta_j,\e ] \, \Big] \,
        \widehat\phi_0^{(j)}(k_0^{(j)}) \, \overline{\widehat\phi_0^{(j)}(k_{8N+1}^{(j)})} \;,
        \label{ExpphinN-2}
\eeqn
where $\e=\frac1t$. The notations are the same as in the
proof of Lemma  {\ref{mainlm2}}. See (\ref{Kj-def-1}) for
the definition of $\widetilde K^{(j)}_{4N,N,\kappa}$.
We note that here, the propagators on each particle line previously
labeled by $n$ and $n+1$ are absent,
since we are considering $\widetilde\phi_{4N,N,\theta_{m-1}}(\theta_m)$
instead of $\phi_{n,N,\theta_{j-m}}(\theta_m)$, see (\ref{eq:tilde-phi-def-1}).

Let $\Pi_{r;8N,4N}^{conn}$ denote the subset of $\Pi_{r;8N,4N}$ of
completely connected graphs.

\begin{lemma}
\label{conngrbd-3}
Let $r\geq1$, $r\in\N$,   and let
$\pi\in\Pi_{r;8N,4N}^{conn}$
be a completely connected graph.
Then, for every $T=\lambda^2\e^{-1}>0$, there exists a finite constant $c=c(T)$ such that
\eqn
        |\amp_\delta(\pi)| \; \leq \; \kappa^{-2rN}
        (\log\frac1\e)^{3r }(c\lambda^2\e^{-1}
        \log\frac1\e)^{\frac{r\bar n}{2}} \;.
\eeqn
\end{lemma}

\prf
We modify the proof of Lemma {\ref{lm:apriori-1}} in the following
manner. We observe that (\ref{ExpphinN-1}) contains $r(6N+2)$
propagators with imaginary parts $\pm i\kappa\e$, and $2rN$ propagators
with imaginary parts $\pm i\e$.
In the proof of Lemma {\ref{lm:apriori-1}}, $4rN$ out of all propagators were
estimated in $L^\infty$, while the rest was estimated in $L^1$.
The fact that there are two propagators less
per reduced 1-particle line leads to an improvement over
the estimates of Lemma {\ref{conngrbd-1}} which we, however, do not need to exploit.

Carrying out the same arguments line by line, we estimate
$4rN$ out of all propagators in (\ref{ExpphinN-2})
in $L^\infty$. There are $2rN$ propagators carrying an imaginary
part $\pm i\e$ in the denominator.
By the pigeonhole principle, at least $2rN$
propagators bounded in $L^\infty$ have a denominator
with an imaginary part $\pm i\kappa\e$. From each of those, one obtains
an improvement of the a priori bound in Lemma {\ref{lm:apriori-1}}
by a factor $\kappa^{-1}$. This is because all propagators estimated in
$L^\infty$ in Lemma {\ref{lm:apriori-1}} were bounded by $\frac1\e$.
In total, one gains a factor of at least $\kappa^{-2rN}$ over the
estimate in Lemma {\ref{lm:apriori-1}}.

For a more detailed exposition of arguments concerning the time partitioning
method, we refer to \cite{erdyau}.
\endprf

\begin{lemma}
\label{Nondiscsum-est3}
Let $r\in2\N$. Then,
for every $T>0$, there exists a finite constant $c=c(T)$ such that
\eqn
        \sum_{\pi\in\Pi_{r;8N,4N}}|\amp_{\delta}(\pi)| \; \leq \;
        (4rN)!\kappa^{-2rN}(\log\frac1\e)^{3r} (c\lambda^2\e^{-1}
        \log\frac1\e)^{4rN} \;.
\eeqn
\end{lemma}

\prf
Let $\pi\in\Pi_{r;8N,4N}$ have $m$ connectivity components,
and let $\pi$ comprise
$s_1,\dots,s_m$ particle lines, where $\sum_{l=1}^m s_l=r$.
Then,
\eqn
        |\amp_{\delta}(\pi)|&\leq&\kappa^{-2N\sum_{l=1}^m s_l}
        (\log\frac1\e)^{3\sum_{l=1}^m  s_l }
        (c\e^{-1}\lambda^2\log\frac1\e)^{\frac{\bar n\sum_{l=1}^m s_l}{2}}
        \nonumber\\
        &\leq&\kappa^{-2rN}
        (\log\frac1\e)^{3r}
        (c\e^{-1}\lambda^2\log\frac1\e)^{4rN}\;.
\eeqn
Moreover, the number of elements of $\Pi_{r;8N,4N}$ is bounded by $(4rN)!\,2^{4rN}$.
\endprf

The corresponding sum over disconnected graphs can be estimated
by the bound in Lemma {\ref{Nondiscsum-est2}}.
This proves Lemma {\ref{mainlm3}}.

\subsection*{Acknowledgements}

I am deeply grateful to H.-T. Yau and L. Erd\"os for their support, encouragement, advice, and generosity.
I have benefitted immensely from numerous discussions with them, in later stages
of this work especially from conversations with L. Erd\"os.
I also thank H.-T. Yau for his very generous hospitality during two visits at Stanford University.
I am most grateful to the anonymous referee for very detailed and helpful comments,
and for pointing out an error related to the WKB initial condition in an earlier version
of the manuscript.
This work was supported by NSF grants DMS-0407644 and DMS-0524909,
partially by a Courant Instructorship, and partially by a grant of
the NYU Research Challenge Fund Program.


\parskip = 0 pt
\parindent = 0 pt

\newpage

\centerline{\epsffile{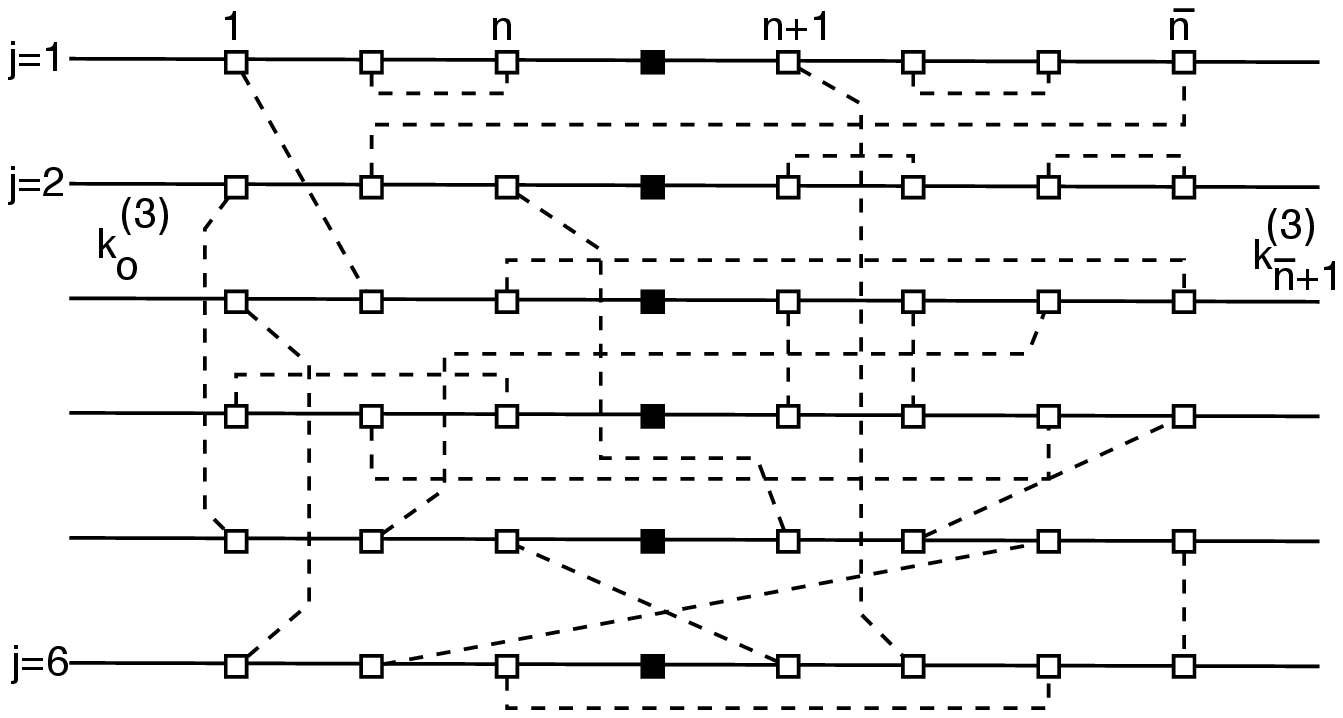} }

\noindent{Figure 1.} A (completely connected) contraction graph for
the case $r=6$, $n=3$, $\bar n=7$. The $\hJe$-vertices are drawn in
black, while the $\widehat V_\omega$-vertices are shown in white.
The $r$ particle lines are solid, while the lines corresponding to
contractions of pairings of random potentials are dashed. For $j=3$
in the notation of (\ref{End-1}), the momenta $k_0^{(3)}$ and
$k_{\bar n+1}^{(3)}$ are written above the corresponding propagator
lines.
\\
\\

\newpage

\centerline{\epsffile{lpc-fig2.epsf} }

\noindent{Figure 2.}
The decomposition of the graph $\pi$ in Figure 1 into reduced 1-particle lines, with
the exception of the particle lines labeled by $j=1$ and $j=2$.
A numbered vertex with label $\ell$ accounts for a reduced transfer delta carrying the transfer
momentum $u_\ell$, and a label $-\ell$ accounts for one carrying a transfer momentum $-u_\ell$. In this
example, unfilled numbered transfer vertices carry transfer momenta used for $L^\infty$-bounds
in (\ref{eq:ampj-bd-aux-3}), while shaded
transfer vertices carry transfer momenta used for $L^1$-bounds.
\\

\pagebreak

\centerline{\epsffile{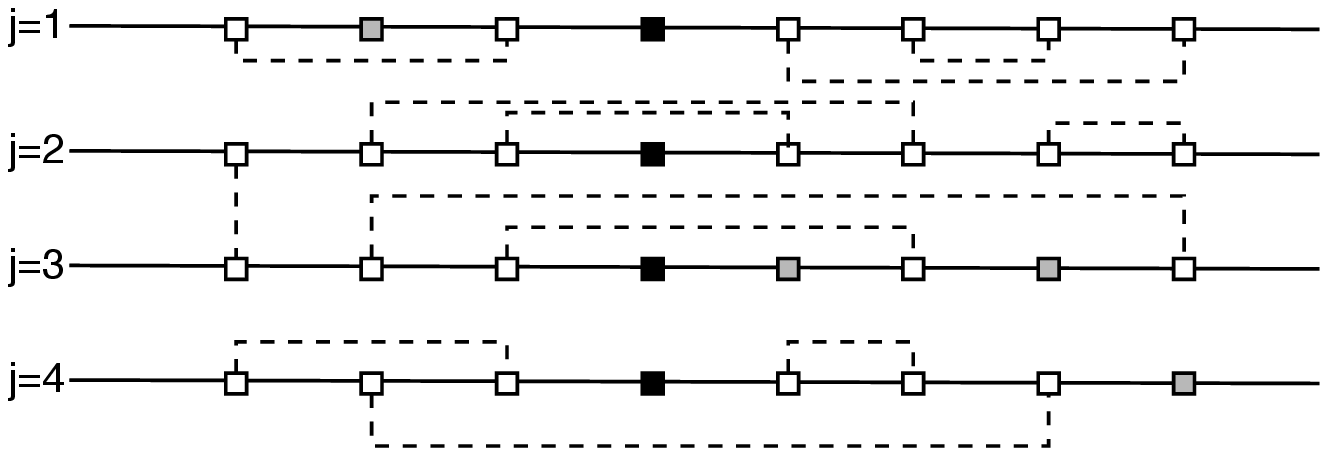} }

\noindent{Figure 3.}
An example unrelated to that in Figures 1 and 2. Here, all reduced transfer vertices are shaded,
while the unreduced transfer vertices and
all internal vertices are unfilled. The reduced 1-particle line with
$j=1$ contains an immediate recollision with a reduced transfer vertex insertion,
and a nesting subgraph.
The reduced 1-particle lines with $j=2,3$ define a ladder diagram  with
two rungs each, and decorated by a immediate recollision,
and connected by a transfer contraction line.
The reduced 1-particle line with $j=4$ contains a crossing subgraph.
\\

\end{document}